\documentclass[aps,pre,preprint,tightenlines,superscriptaddress,longbibliography]{revtex4-1}

\usepackage{amsmath}
\usepackage{amssymb}
\usepackage{hyperref}
\usepackage[arrow, matrix, curve]{xy}
\usepackage{bm}
\usepackage{yhmath}
\usepackage{color}

\usepackage{hyperref}

\usepackage{amsmath,amssymb,graphicx,textcomp}
\usepackage{cleveref} 
\usepackage{verbatim}

\DeclareMathOperator\Real{Re}

\newcommand\img{\mathsf{i}}
\newcommand\peRe{\text{Pe}}

\usepackage{color}
\newcommand\diff{\mathrm{d}}

\usepackage[usenames,dvipsnames]{xcolor}
\hypersetup{colorlinks=true, linkcolor=BrickRed, urlcolor=blue!50!black, citecolor=blue!50!black}

\usepackage[normalem]{ulem}

\makeatletter
\newcommand\hide@visible[1]{%
  \bgroup\fboxsep=.3ex\colorbox{Gray}{begin hide}%
  #1\colorbox{Gray}{end hide}\egroup%
}
\newcommand\hide@hidden[1]{%
  \bgroup\fboxsep=.3ex\colorbox{Gray}{hidden text}%
}
\newcommand\hide@invisible[1]{}
\newcommand\makevisible{\let\hide\hide@visible}
\newcommand\makehidden{\let\hide\hide@hidden}
\newcommand\makeinvisible{\let\hide\hide@invisible}
\makeatother
\makehidden

\begin{document}

\texttt{published in Physical Review Fluids \textbf{3}, 103301 (2018)}


\title{Time-dependent active microrheology in dilute colloidal suspensions}


\author{Sebastian Leitmann}
\affiliation{Institut f\"ur Theoretische Physik, Universit\"at Innsbruck, Technikerstra{\ss}e~21A,  A-6020 Innsbruck, Austria}

\author{Suvendu Mandal}
\affiliation{Institut f\"ur Theoretische Physik, Universit\"at Innsbruck, Technikerstra{\ss}e~21A,  A-6020 Innsbruck, Austria}

\author{Matthias Fuchs}
\affiliation{Fachbereich Physik, Universit\"at Konstanz, 78457 Konstanz, Germany}

\author{Antonio M. Puertas}
\affiliation{Group of Complex Fluids Physics, Department of Applied Physics, University of Almer\'{\i}a, 04120 Almer\'{\i}a, Spain}

\author{Thomas Franosch}
\affiliation{Institut f\"ur Theoretische Physik, Universit\"at Innsbruck, Technikerstra{\ss}e~21A,  A-6020 Innsbruck, Austria}
\email[]{thomas.franosch@uibk.ac.at}

 
\date{\today}




\begin{abstract}
In a microrheological set-up a single probe particle immersed in a complex fluid is exposed to a strong external force
driving the system out of equilibrium. Here, we elaborate analytically the time-dependent response of a probe particle
in a dilute suspension of Brownian particles to a large step-force, exact in first order of the density of the bath
particles. The time-dependent drift velocity approaches its stationary state value exponentially fast for arbitrarily small
driving in striking contrast to the power-law prediction of linear response encoded in the long-time tails of the
velocity autocorrelation function. We show that the stationary-state behavior depends nonanalytically on the driving
force and connect this behavior to the persistent correlations in the equilibrium state. We argue that this relation
holds generically. Furthermore, we elaborate that the fluctuations in the direction of the force display transient
superdiffusive behavior.


\end{abstract}

\maketitle


\section{Introduction} 

While the static and dynamic properties of interacting many-particle systems in equilibrium encode the linear response
via the fluctuation-dissipation theorem, corresponding principles generally applicable for systems driven far from
equilibrium remain a grand challenge in statistical physics.  Soft matter systems are ideally suited to study such
nonequilibrium phenomena, since their defining characteristic is that they are strongly susceptible to forces.
Conceptually, the simplest experimentally realizable system then consists of an interacting colloidal suspension that is
driven out of equilibrium by a strong external force acting on a single probe particle. This setup constitutes the basic
paradigm for active microrheology with the principle goal to infer material properties beyond the linear
regime~\cite{Squires:PoF_17:2005,Khair:JFM_557:2006,Wilson:JPCB_113:2009,Zia:JFM_658:2010,Sriram:PoF_22:2010,Wilson:PCCP_13:2011,Zia:JoR_56:2012,Swan:PoF_25:2013,Puertas:JPCM_26:2014,Zia:Springer:2015,Furst:Oxford:2017}. 

The nonlinear mobility in the stationary state $\mu(\text{Pe})$ for a suspension of interacting Brownian particles has
been derived to first order of the density of bath particles $n$ in the seminal work by Squires and
Brady~\cite{Squires:PoF_17:2005} in terms of an asymptotic expansion  
\begin{align} \label{eq:series_expansion_squires}
\mu(\text{Pe})/\mu = 1 - \frac{2\pi n \sigma^3}{3} \frac{D_a}{D_r} \biggl[1 - \frac{2}{15}\text{Pe}^2 + \frac{1}{8}|\text{Pe}|^3 -
\frac{128}{1575}\text{Pe}^4 + \mathcal{O}(|\text{Pe}|^5)\biggr], \quad \text{Pe} \to 0.
\end{align}
where $\mu$ is the mobility of the probe particle at infinite dilution with diffusion coefficient $D_a$, and the
P{\'e}clet number $\text{Pe}$ is a suitable dimensionless measure for the driving force (see below). The motion of the
bath particles with diffusion coefficient $D_b$ enters in terms of the diffusion coefficient of the relative motion,
$D_r = D_a + D_b$.  

The leading term of the corresponding asymptotic expansion of the fluctuations around the average
drift has also been achieved~\cite{Zia:JFM_658:2010}, revealing a long-time diffusion coefficient that can become
arbitrarily large for strong driving. Active microrheology has found fruitful applications in dense colloidal systems in
the vicinity of the glass transition, in particular, computer simulations have revealed (transient) superdiffusion and
enhanced diffusivities~\cite{Winter:PRL_108:2012,Winter:JCP_138:2013}. Certain phenomena in the nonlinear regime have
also been rationalized within a mode-coupling
approach~\cite{Gazuz:PRL_102:2009,Gnann:SM_7:2011,Gnann:PRE_86:2012,Harrer:JPCM_24:2012,Gazuz:PRE_87:2013,Wang:PRE_89:2014,Gruber:PRE_94:2016},
continuous-time random walks~\cite{Jack:PRE_78:2008,Schroer:PRL_110:2013,Schroer:JCP_138:2013,Burioni:CTP_62:2014},
Langevin equations~\cite{Demery:NJP_16:2014,Demery:PRE_91:2015}, and kinetic theory~\cite{Wang:PRE_93:2016}.

Exact results beyond the stationary state have been accomplished for driven transport in lattice models, e.g., for a
biased intruder in a dense crowded environment of mobile hard-core
obstacles~\cite{Benichou:PRL_111:2013,Illien:PRL_111:2013,Benichou:JSMTE_05:2013,Benichou:PRL_113:2014,Illien:PRL_113:2014,Illien:JSMTE:2015}
and the complementary limit of a tracer in a dilute quenched array of obstacles~\cite{Leitmann:PRL_111:2013,
Leitmann:PRL_118:2017,Leitmann:JPhysA:2018}. Due to repeated encounters with the same obstacle one finds that the nonlinear
force-dependent mobility in the stationary state becomes a nonanalytic function of the driving force.

For the case of colloids in continuum, one infers that the mobility also becomes a nonanalytic function in the
P{\'e}clet number, signaled by $|\text{Pe}|^3$ in the asymptotic
expansion~[Eq.~\eqref{eq:series_expansion_squires}], since the mobility has to be an even function of $\text{Pe}$.

The frequency-dependent linear mobility $\hat{\mu}(\omega)$ can be obtained via the fluctuation-dissipation theorem from the
velocity-autocorrelation function in equilibrium, which has been calculated earlier to the same order in the packing 
fraction~\cite{Hanna:PhysA_111:1982,Hanna:JPAMG_14:1982,Ackerson:JCP_76:1982,Felderhof:PhysA_122:1983,Khair:JoR_49:2005}
\begin{align} \label{eq:mobility_equilibrium}
-\img\omega\hat{\mu}(\omega)/\mu = 1 - \frac{2\pi n \sigma^3}{3}\frac{D_a}{D_r}\frac{1 + \sqrt{-\img\omega\tau}}{1 + \sqrt{-\img\omega\tau} - \img\omega\tau/2},
\end{align}
with diffusive time scale $\tau = \sigma^2/D_r$.
Here, the nonanalytic contribution in the frequency $\omega$ reflects the well-known long-time tails in the
velocity-autocorrelation function $Z(t)\simeq -A t^{-5/2}$, $A > 0$, also familiar from the Lorentz
model~\cite{Ernst:PLA_34:1971,vanBeijeren:RMP_54:1982,Nieuwenhuizen:PRL_57:1986,Hoefling:PRL_98:2007,Franosch:CP_375:2010,Bauer:EPJ_189:2010}.

These results immediately raise the question how the nonanalytic contributions in the nonlinear mobility emerge from a
perturbative scheme and whether they are related to the persistent correlations in equilibrium? 
How fast
is the nonequilibrium steady state approached from an initial equilibrium state in comparison to the predictions of
linear response? What is the intermediate time-dependent behavior of the fluctuations connecting the short-time
motion to the drastically enhanced long-time diffusion? These questions will be answered by 
solving the two-particle Smoluchowski equation for the 
time-dependent dynamics in the presence of a strong force for the first time. Thereby, we
unify the previous approaches for the driven stationary state~\cite{Squires:PoF_17:2005,Zia:JFM_658:2010} and the
time-dependent equilibrium dynamics~\cite{Hanna:PhysA_111:1982,Hanna:JPAMG_14:1982,Ackerson:JCP_76:1982,Felderhof:PhysA_122:1983} and reveal the interplay between persistent correlations and nonequilibrium driving.
The solution enables us to fully address the time-dependent approach to the stationary state, in principle for all moments
of the displacement along the force. The two lowest moments, the mobility and the fluctuations, will be
elaborated and compared to computer simulations. 

This work is organized as follows. In Sec.~\ref{sec:model_and_solution_strategy} the underlying model is introduced
followed by the complete solution strategy. The main result of this section is the self-energy encoding the
dynamics between probe and the bath particles. Readers who are primarily concerned about the results may skip this
section and jump directly to Sec.~\ref{sec:cumulants_of_displacement} where we determine the time-dependent behavior
in terms of the mobility and the fluctuations along the applied force. The obtained results from the analytic solution
are compared to computer simulations and the phenomena involved in the transition from the initial equilibrium state
to the new stationary state are discussed. In Sec.~\ref{sec:summary_and_conclusion}, the key results of this work are
summarized followed by general conclusions.

\section{Model and solution strategy} \label{sec:model_and_solution_strategy}

We solve for the time-dependent dynamics of a probe particle pulled by a force in the presence of other bath particles.
In first order of the density of bath particles, the dynamics is completely encoded by the interactions of the probe
particle with a single bath particle~\cite{Felderhof:PhysA_121:1983,Squires:PoF_17:2005}. Discarding inertial effects it is sufficient to consider the two-particle Smoluchowski equation for
probe and bath particle. The problem can be expressed via the independent motion of the center of
diffusion and the relative distance between both particles. The dynamics of the relative distance is described in the
frequency domain in terms of a self-energy which encodes all corrections to the dynamics of the probe
particle due to interactions.

\subsection{Two-particle Smoluchowski equation}
We consider a single probe particle $a$ interacting with a bath particle $b$ with
bare diffusion coefficients $D_a$ and $D_b$, respectively. The particles interact by mutual hard-core exclusion with exclusion distance
$\sigma$. At time $t = 0$ an external constant force $\vec{F}$ is switched on, pulling the probe particle and driving
the system from its initial equilibrium state into a nonequilibrium stationary state. We describe the state of the
system by the conditional probability density $\Psi(\vec{r}_a, \vec{r}_b, t| \vec{r}_a^{\,\prime}, \vec{r}_b^{\,\prime})$
for probe particle $a$ and bath particle $b$ to be at positions $\vec{r}_a$, $\vec{r}_b$ at time $t$ provided they start
at initial positions
$\vec{r}_a^{\,\prime}$, $\vec{r}_b^{\,\prime}$ at time $t=0$. 
Since the system is initially in equilibrium, the corresponding initial condition reads $\Psi(\vec{r}_a, \vec{r}_b, t =
0| \vec{r}_a^{\,\prime}, \vec{r}_b^{\,\prime}) = \vartheta(|\vec{r}_a^{\,\prime}-\vec{r}_b^{\,\prime}| -
\sigma)\delta(\vec{r}_a - \vec{r}_a^{\,\prime})\delta(\vec{r}_b - \vec{r}_b^{\,\prime})/V$, where the Heaviside-function
$\vartheta(\cdot)$ accounts for the mutual exclusion. 
The limit of large box sizes $V\to\infty$ is anticipated throughout. At time $t=0$, the force $\vec{F}$ is switched on and the
probability density evolves according to the Smoluchowski equation
\begin{align}
 \partial_t \Psi = (D_a \nabla_a^2+D_b\nabla_b^2) \Psi - \mu \vec{F} \cdot \vec{\nabla}_a \Psi,
\end{align}
with bare mobility $\mu = D_a/k_\text{B} T$ of the probe particle and the thermal scale $k_\text{B}
T$. The hard-core interaction between probe and bath particle is encoded in the no-flux boundary condition
\begin{align}
(\vec{r}_a-\vec{r}_b)\cdot  \Bigl[ \mu \vec{F} \Psi - (D_a\vec{\nabla}_a-D_b\vec{\nabla}_b) \Psi \Bigr]= 0, \quad
\text{for } |\vec{r}_a-\vec{r}_b| = \sigma. 
\end{align}

We introduce new coordinates for the center of diffusion $\vec{R} = (D_b\vec{r}_a + D_a\vec{r}_b)/(D_a + D_b)$, and
the relative distance $\vec{r} = \vec{r}_a - \vec{r}_b$. After transformation, the Smoluchowski equation in these
adapted coordinates reads
\begin{align}
\begin{split} \label{eq:smoluchowski_equation_R_r}
 \partial_t \Psi &= \frac{D_a D_b}{D_r} \nabla^2_R \Psi -\frac{\mu D_b}{D_r}  \vec{F} \cdot \vec{\nabla}_R \Psi +
D_r\nabla^2_r \Psi - \mu \vec{F} \cdot \vec{\nabla}_r \Psi, 
\end{split}
\end{align}
with the diffusion coefficient for the relative motion, $D_r = D_a + D_b$.
Similarly, the no-flux boundary condition transforms to
\begin{align}
\begin{split} \label{eq:smoluchowski_equation_R_r_boundary}
\vec{r} \cdot \left[ \mu \vec{F} \Psi - D_r \vec{\nabla}_r \Psi \right] = 0, \quad \text{for } |\vec{r}| = \sigma.
\end{split}
\end{align}
The Smoluchowski equation and the no-flux boundary condition in the new coordinates
[Eq.~\eqref{eq:smoluchowski_equation_R_r} and~\eqref{eq:smoluchowski_equation_R_r_boundary}] reveal that the motion of
the center of diffusion $\vec{R}$ and the dynamics of the relative motion $\vec{r}$, are independent. Thus, the
conditional probability factorizes into a simple Gaussian with diffusion coefficient $D_a D_b/D_r$ and drift $\mu
\vec{F} D_b/D_r$ for the center of diffusion $\vec{R}$, and the conditional probability $\psi(\vec{r}, t| \vec{r}\,')$ to find relative
distances $\vec{r}$ and $\vec{r}\,'$ at times $t$ and $0$. It fulfills the reduced Smoluchowski equation 
\begin{align}\label{eq:smoluchowski_equation_r}
 \partial_t \psi = D_r \nabla^2_r \psi - \mu \vec{F} \cdot \vec{\nabla}_r \psi ,
\end{align}
and the initial condition is provided by the equilibrium state $\psi(\vec{r}, t = 0| \vec{r}\,') = \vartheta(|\vec{r}\,'| -
\sigma)\delta(\vec{r}-\vec{r}\,')$. Similarly, the no-flux boundary condition for the conditional probability $\psi$
reads
\begin{align} \label{eq:boundary_condition_r}
\vec{r} \cdot \left[ \mu \vec{F} \psi - D_r \vec{\nabla}_r \psi \right] = 0, \qquad \text{for } |\vec{r}| = \sigma.
\end{align}

It is natural to measure the force in terms of the dimensionless P{\'e}clet number $\peRe = \mu F \sigma/D_r$, that
already appeared in Eq.~\eqref{eq:series_expansion_squires}~\cite{Squires:PoF_17:2005,Zia:JFM_658:2010},
which can be written also as $\peRe = (F \sigma / k_\text{B} T) D_a/ D_r$ by using the Stokes-Einstein relation.
This definition encompasses the case of equal sized colloids, $\text{Pe} = F \sigma/2 k_\text{B} T$, as well as the
Lorentz model, $\text{Pe} = F \sigma/k_\text{B} T$, where the bath particles are fixed in space ($D_b = 0$).

Our main quantity of interest is the intermediate scattering function $\langle
e^{-\img\vec{q}\cdot\Delta\vec{r}_a(t)}\rangle$ for the displacement $\Delta \vec{r}_a(t) = \vec{r}_a(t) - \vec{r}_a(0)$
of the probe particle, from which in principle all moments of the displacement can be extracted by derivatives with
respect to the wave vector $\vec{q}$. 
In new coordinates, $\Delta \vec{r}_a(t) = \Delta \vec{R}(t) + D_a
\Delta \vec{r}(t)/D_r$, the intermediate scattering function can be expressed as
\begin{align} \label{eq:int_scat_prod}
\langle e^{-\img \vec{q}\cdot \Delta \vec{r}_a(t)}\rangle = \langle e^{-\img \vec{q} \cdot\Delta \vec{R}(t)}\rangle \langle
e^{-\img (D_a/D_r) \vec{q} \cdot \Delta \vec{r}(t)}\rangle, 
\end{align}
where we used the independence of $\vec{R}$ and $\vec{r}$~[Eq.~\eqref{eq:smoluchowski_equation_R_r}].
Since the dynamics of the center of diffusion is Gaussian with diffusion coefficient $D_a D_b /D_r$ and drift
$\mu \vec{F} D_b/D_r$, one finds immediately
\begin{align} \label{eq:int_scat_gaussian}
\langle e^{-\img \vec{q}\cdot \Delta \vec{R}(t)}\rangle = e^{-\img \vec{q}\cdot (\mu \vec{F} D_b /D_r) t}
e^{-q^2 (D_a D_b/D_r) t} .
\end{align}
The remaining task is the calculation of the intermediate scattering function $\langle e^{-\img (D_a/D_r)
\vec{q}\cdot\Delta \vec{r}(t)}\rangle$ for the relative motion $\Delta \vec{r}(t)$ between probe and bath particle.

\subsection{Dynamics of the relative motion}

This part contains the detailed calculation of the relative motion and contains the heart of our analytic approach.
The solution strategy is adapted from Felderhof's approach for the equilibrium
dynamics~\cite{Felderhof:PhysA_122:1983}.

The intermediate scattering function, viz. the characteristic function of the relative
displacement $\Delta \vec{r}(t) = \vec{r}(t) - \vec{r}(0)$,
\begin{align} \label{eq:intermediate_scattering_function_r}
\langle e^{-\img\vec{q}\cdot\Delta \vec{r}(t)}\rangle = \int\!\diff^3 r \int\! \frac{\diff^3 r'}{V}\ e^{-\img\vec{q}\cdot(\vec{r} -
\vec{r}\,')}\psi(\vec{r}, t | \vec{r}\,'),
\end{align}
allows extracting the moments by series expansion in the wave vector $\vec{q}$. 
To make analytic progress, we only consider the axially symmetric case, where the wave vector and force are aligned
$\vec{q}\parallel\vec{F}$. This enables us to determine the motion in the direction of the force, the time-dependent motion
perpendicular to the force is left for future analysis. We perform a temporal Fourier-Laplace transform and a spatial
Fourier transform, 
\begin{align}
\hat{\psi}_{q}(\vec{r},\omega) := \int_0^\infty\!\diff t\, e^{\img\omega t}
\int\!\frac{\diff^3 r'}{\sqrt{V}}\, e^{\img\vec{q}\cdot\vec{r}\,'} \psi(\vec{r},t|\vec{r}\,') .
\end{align} 
The transformed quantity $\hat{\psi}_{q}(\vec{r},\omega)$ is connected to the propagator $G(q,\omega)$ via a spatial
Fourier transform $G(q,\omega) := \int\diff^3 r \ e^{-\img\vec{q}\cdot\vec{r}} \hat{\psi}_{q}(\vec{r},\omega)/\sqrt{V}$.
From Eq.~\eqref{eq:smoluchowski_equation_r} and the initial condition one derives the equation of motion~\cite{Felderhof:PhysA_122:1983}
\begin{align} \label{eq:motion_conditional_prob}
 (-\img \omega - D_r\nabla^2 +\mu \vec{F} \cdot \vec{\nabla} ) \hat{\psi}_{q} =
\frac{1}{\sqrt{V}} e^{\img \vec{q} \cdot \vec{r} } \vartheta(|\vec{r}| - \sigma).
\end{align}
In particular, $\hat{\psi}_{q}(\vec{r},\omega) = 0$ for $|\vec{r}| < \sigma$, reflects the hard-core exclusion.
Without interaction, we denote the respective conditional probability by $\hat{\psi}_{q}^0$ which evolves according to
\begin{align} \label{eq:motion_conditional_prob_free}
 (-\img \omega - D_r\nabla^2 +\mu \vec{F} \cdot \vec{\nabla} ) \hat{\psi}_{q}^0 =
\frac{1}{\sqrt{V}} e^{\img \vec{q} \cdot \vec{r} }.
\end{align}
The free motion allows for a plane-wave solution of the form
\begin{align} \label{eq:conditional_prob_free_sol}
\hat{\psi}_{q}^0(\vec{r}, \omega) = \frac{e^{\img \vec{q} \cdot \vec{r}}/\sqrt{V} }{-\img \omega +
D_r q^2 + \img\mu F q} ,
\end{align}
which is valid for all $\vec{r} \in V$ and the respective free propagator can be read off as
\begin{align}
G_0(q,\omega) = \frac{1}{-\img\omega + D_r q^2 + \img \mu F q}.
\end{align}
In particular, the plane-wave solution can be written as $\hat{\psi}_q^0(\vec{r},\omega) = G_0(q,\omega)
e^{\img\vec{q}\cdot\vec{r}}/\sqrt{V}$.
To make further progress, we observe that the difference of Eqs. \eqref{eq:motion_conditional_prob} and
\eqref{eq:motion_conditional_prob_free},
\begin{align}\label{eq:Helmholtz}
 (-\img \omega - D_r \nabla^2 +\mu \vec{F} \cdot \vec{\nabla} ) [\hat{\psi}_{q}(\vec{r},\omega) -
\hat{\psi}_{q}^0(\vec{r},\omega)]=
0,\quad\text{for }|\vec{r}|>\sigma,
\end{align}
vanishes for terminal distance $|\vec{r}|$ larger than the exclusion distance $\sigma$.
We are interested in the forward-scattering amplitude 
\begin{align}
G(q,\omega) - G_0(q,\omega) = \frac{1}{\sqrt{V}} \int \diff^3 r\  e^{-\img \vec{q} \cdot \vec{r}} [\hat{\psi}_{q}(\vec{r},\omega) -
\hat{\psi}_{q}^0(\vec{r},\omega)] 
\end{align}
which we separate into contributions outside and inside of the region of overlap of probe and bath particle:
\begin{align}\label{eq:Fourier_integral}
\begin{split}
G(q,\omega) - G_0(q,\omega) 
&= \frac{1}{\sqrt{V}} \int_{|\vec{r}|> \sigma} \diff^3 r\  e^{-\img \vec{q} \cdot \vec{r}} [\hat{\psi}_{q}(\vec{r},\omega) -
\hat{\psi}_{q}^0(\vec{r},\omega)] + \\
&\phantom{=}\ + \frac{1}{\sqrt{V}} \int_{|\vec{r}|< \sigma} \diff^3 r\  e^{-\img \vec{q} \cdot \vec{r}}  [\hat{\psi}_{q}(\vec{r},\omega) -
\hat{\psi}_{q}^0(\vec{r},\omega)].
\end{split}
\end{align}
In the second term we observe that the conditional probability $\hat{\psi}_{\vec{q}}$ vanishes for
terminal positions inside the obstacle. Hence, there we can immediately compute the integral
\begin{align}
 \frac{1}{\sqrt{V} } \int_{|\vec{r}| < \sigma }\diff^3 r\ e^{-\img \vec{q} \cdot \vec{r}}
\hat{\psi}_{q}^0(\vec{r},\omega) &= \frac{4\pi \sigma^3/3 V }{-\img \omega + D_r q^2 + \img\mu F q} . 
\end{align}

To calculate the first term in Eq.~\eqref{eq:Fourier_integral}, it is advantageous to convert the volume integral
into a surface integral relying on the equations of motion. We use the auxiliary variable $\chi(\vec{r}) =
\hat{\psi}_{q}(\vec{r},\omega) - \hat{\psi}_{q}^0(\vec{r},\omega)$ and investigate the identity
\begin{align}
\int_{|\vec{r}| > \sigma } \diff^3 r\ e^{-\img \vec{q}\cdot \vec{r} } (-\img \omega - D_r \nabla^2 +\mu
\vec{F} \cdot \vec{\nabla} ) \chi(\vec{r}) = 0,
\end{align}
which follows from Eq.~\eqref{eq:Helmholtz}.
Then, for the drift, we obtain 
\begin{align}
\begin{split}
\int_{|\vec{r}| > \sigma}\diff^3 r\ e^{-\img \vec{q}\cdot \vec{r} } (-\mu \vec{F} \cdot \vec{\nabla} ) \chi(\vec{r}) 
&= \int_{|\vec{r}| > \sigma}\diff^3 r\ (-\mu \vec{F} \cdot \vec{\nabla} ) \Bigl[ e^{-\img \vec{q}\cdot \vec{r} } \chi(\vec{r}) \Bigr] +
\int_{|\vec{r}| > \sigma}\diff^3 r\ \chi(\vec{r})  \mu \vec{F} \cdot \vec{\nabla}   e^{-\img \vec{q}\cdot \vec{r} }  \\
&=\int \diff S_{\vec{r}} \,    \hat{\vec{r}} \cdot \mu \vec{F}   e^{-\img \vec{q}\cdot \vec{r} }  \chi(\vec{r})  +
\int_{|\vec{r}| > \sigma}\diff^3 r\ \chi(\vec{r})  \mu \vec{F} \cdot \vec{\nabla}   e^{-\img \vec{q}\cdot \vec{r} } ,
\end{split}
\end{align}
where we introduced the unit vector $\hat{\vec{r}} = \vec{r}/|\vec{r}|$ and the surface element $\diff S_{\vec{r}}$ of
the sphere of radius $\sigma$. Similarly, for the diffusive contribution, we derive
\begin{align}
\begin{split}
\int_{|\vec{r}| > \sigma}\diff^3 r\ e^{-\img \vec{q}\cdot \vec{r} } D_r \nabla^2 \chi(\vec{r}) 
&=\int_{|\vec{r}| > \sigma}\diff^3 r\ D_r \vec{\nabla}\cdot  \bigl[ e^{-\img \vec{q}\cdot \vec{r} } \vec{\nabla} \chi(\vec{r}) \bigr] 
- \int_{|\vec{r}| > \sigma}\diff^3 r\ \bigl[ D_r\vec{\nabla}\chi(\vec{r}) \bigr] \cdot \vec{\nabla} e^{-\img \vec{q}\cdot \vec{r}} \\ 
&=-\int \diff S_{\vec{r}} \, D_r \hat{\vec{r}} \cdot e^{-\img \vec{q}\cdot \vec{r}} \vec{\nabla} \chi(\vec{r}) 
- \int_{|\vec{r}| > \sigma}\diff^3 r\ D_r \vec{\nabla} \cdot \bigl[\chi(\vec{r})\ \vec{\nabla} e^{-\img \vec{q}\cdot
\vec{r}}\bigr]\ + \\
&\phantom{=}\ + \int_{|\vec{r}| > \sigma}\diff^3 r\ \chi(\vec{r}) D_r \nabla^2  e^{-\img \vec{q}\cdot \vec{r} } \\
&=- \int \diff S_{\vec{r}} \, D_r \hat{\vec{r}} \cdot e^{-\img \vec{q}\cdot \vec{r}} \vec{\nabla} \chi(\vec{r})  +
\int \diff S_{\vec{r}}  \, D_r  \hat{\vec{r}} \cdot \bigl[\chi(\vec{r})\ \vec{\nabla} e^{-\img \vec{q}\cdot \vec{r}}\bigr]\ + \\
&\phantom{=}\  + \int_{|\vec{r}| > \sigma}\diff^3 r\ \chi(\vec{r}) D_r \nabla^2  e^{-\img \vec{q}\cdot \vec{r} } .
\end{split}
\end{align}
Together, this yields
\begin{align}
\begin{split}
(-\img \omega + D_r q^2 + \img \mu \vec{F} \cdot \vec{q} ) \int_{|\vec{r}| > \sigma}\diff^3 \vec{r}\ \chi(\vec{r})
e^{-\img \vec{q}\cdot \vec{r} } 
&= \int \diff S_{\vec{r}} \, \hat{\vec{r}} \cdot \bigl[ \mu \vec{F} \chi(\vec{r}) - D_r \vec{\nabla} \chi(\vec{r})
\bigr] e^{-\img \vec{q} \cdot \vec{r} }\ + \\ 
&\phantom{=}\ + \int \diff S_{\vec{r}} \, D_r \hat{\vec{r}} \cdot \bigl[\chi(\vec{r})\ \vec{\nabla} e^{-\img \vec{q}\cdot \vec{r} }\bigr] .
\end{split}
\end{align}
The first term on the right-hand side simplifies by the no-flux boundary condition~[Eq.~\eqref{eq:boundary_condition_r}] and the plane wave solution
$\hat{\psi}_{q}^0$~[Eq.~\eqref{eq:conditional_prob_free_sol}]:
\begin{align}
\begin{split}
\int \diff S_{\vec{r}} \,  \hat{\vec{r}} \cdot \left[ \mu \vec{F} \chi - D_r \vec{\nabla} \chi \right] e^{-\img \vec{q} \cdot \vec{r} }   
&= -\int \diff S_{\vec{r}} \,  e^{-\img \vec{q} \cdot \vec{r} }   \hat{\vec{r}} \cdot \left[ \mu \vec{F}  - D_r
\vec{\nabla}  \right] \hat{\psi}_{\vec{q}}^0(\vec{r},\omega) \\
&= -  \int \diff S_{\vec{r}} \,  e^{-\img \vec{q} \cdot \vec{r} }   \hat{\vec{r}} \cdot \left[ \mu \vec{F}  -
\img D_r \vec{q}   \right]  \hat{\psi}_{\vec{q}}^0(\vec{r},\omega)  \\
&=   \frac{- 1/\sqrt{V}}{-\img\omega + D_r q^2 + \img \mu \vec{F} \cdot \vec{q} } \int \diff S_{\vec{r}}
\,   \hat{\vec{r}} \cdot \left[ \mu \vec{F}  - \img D_r \vec{q}   \right] = 0,
\end{split}
\end{align}
where in the last line the flux integral of a constant vector vanishes.
Collecting results, we obtain an expression for the forward scattering amplitude in terms of a surface integral of the difference of the
conditional probabilities:
\begin{align} 
\begin{split}
\label{eq:forward_scattering_amplitude}
G(q,\omega) -G_0(q,\omega) &= 
G_0(q,\omega)\biggl\{-\frac{4\pi\sigma^3}{3V} -\frac{1}{\sqrt{V}} \int \diff S_{\vec{r}} \, \img D_r  \vec{q} \cdot \hat{\vec{r}}  e^{-\img \vec{q}\cdot \vec{r} } 
\bigl[\hat{\psi}_{q}(\vec{r},\omega) - \hat{\psi}_{q}^0 (\vec{r},\omega) \bigr] \biggr\} .
\end{split}
\end{align}

In order to determine the conditional probabilities, we return to Eq.~\eqref{eq:Helmholtz}.
For the solution of the homogeneous part of Eq.~\eqref{eq:Helmholtz}, we use the imaginary ``gauge transformation'' $\hat{X}(\vec{r},\omega) =
[\hat{\psi}_{q}(\vec{r},\omega) - \hat{\psi}_{q}^0(\vec{r},\omega)] e^{-\mu \vec{F} \cdot \vec{r} /2 D_r}$ such that the
drift term is absorbed in the Laplacian by completing the square, $\vec{\nabla} \to \vec{\nabla} - \mu
\vec{F}/2D_r$~\cite{Nelson:PRE_58:1998,Squires:PoF_17:2005}. This step will have crucial implications for the results at
finite forces. 
Then, the new quantity $\hat{X}$ fulfills the 3d source-free Helmholtz equation 
\begin{align} \label{eq:Helmholtz_equation}
(\kappa^2 - \nabla^2)\hat{X}(\vec{r},\omega) = 0,\quad\text{for }|\vec{r}|>\sigma,
\end{align}
with complex wavenumber $\kappa^2\sigma^2 = -\img \omega\tau+ (\text{Pe}/2)^2$ and diffusive time scale $\tau =
\sigma^2/D_r$. We write the general axially symmetric solution in the form
\begin{align} \label{eq:solution_Helmholtz_equation}
\hat{\psi}_{q}(\vec{r},\omega)-\hat{\psi}^0_{q}(\vec{r},\omega) 
= \frac{ e^{\mu \vec{F}\cdot\vec{r}/2 D_r}/\sqrt{V}}{-\img \omega + D_r q^2 + \img \mu F q}  \sum_{\ell=0}^\infty a_\ell \frac{\text{k}_\ell(\kappa r)}{\text{k}_\ell(\kappa \sigma)} \text{P}_\ell(\cos \vartheta), 
\end{align}
for $|\vec{r}| > \sigma$, with modified spherical Bessel functions of the second kind $\text{k}_\ell(\cdot)$ and
Legendre polynomials $\text{P}_\ell(\cdot)$. 

The expansion coefficients $a_\ell$ for the different angular channels
$\ell$ have to be determined via the no-flux boundary condition~[Eq.~\eqref{eq:boundary_condition_r}]:
\begin{align}
\begin{split}
\left[ \mu  F \cos\vartheta - D_r \frac{\partial }{\partial r} \right] (\hat{\psi}_q - \hat{\psi}_q^0)\biggr|_{r = \sigma} &= -
\left[ \mu  F \cos\vartheta - D_r \frac{\partial }{\partial r} \right] \hat{\psi}_q^0\biggr|_{r = \sigma} .
\end{split}
\end{align}
With the explicit solutions [Eqs.~\eqref{eq:conditional_prob_free_sol} and \eqref{eq:solution_Helmholtz_equation}],
we find that the expansion coefficients $a_\ell$ have to fulfill
 \begin{align}
 \left[ \text{Pe}\,\eta - \sigma \frac{\partial }{\partial r} \right] e^{\text{Pe}\,\eta r /2\sigma} \sum_{\ell=0}^\infty a_\ell \frac{\text{k}_\ell(\kappa r)}{\text{k}_\ell(\kappa \sigma)} \text{P}_\ell(\eta)\biggr|_{r = \sigma} =& -
 \left[ \text{Pe}\, \eta -  \sigma \frac{\partial }{\partial r} \right] e^{\img q \eta r} \biggr|_{r=\sigma} ,
\end{align}
where we abbreviated $\eta = \cos(\vartheta)$ and used the P{\'e}clet number $\text{Pe} = \mu F \sigma/D_r$.
Performing the derivatives the relation can be expressed as
\begin{align}
\begin{split}
e^{-\text{Pe}\,\eta/2} (\text{Pe} - \img q \sigma) \eta e^{\img q \sigma \eta}  
&= \sum_{\ell=0}^\infty a_\ell \biggl[\frac{\kappa \sigma \text{k}_\ell'(\kappa \sigma)}{\text{k}_\ell(\kappa \sigma)}
-  \frac{\text{Pe}}{2} \eta \biggr] \text{P}_\ell(\eta)  \\
&= \sum_{\ell=0}^\infty \biggl[\frac{\kappa \sigma \text{k}_\ell'(\kappa \sigma)}{\text{k}_\ell(\kappa \sigma)} a_\ell \text{P}_\ell(\eta) 
- \frac{\text{Pe}}{2} a_\ell \biggl(\frac{\ell+1}{2\ell +1} \text{P}_{\ell+1}(\eta) + \frac{\ell}{2\ell + 1}\text{P}_{\ell-1}(\eta) \biggr)\biggr] \\
&= \sum_{\ell=0}^\infty \biggl[\frac{\kappa \sigma \text{k}_\ell'(\kappa \sigma)}{\text{k}_\ell(\kappa \sigma)} a_\ell 
- \frac{\text{Pe}}{2} \frac{\ell}{2\ell-1} a_{\ell-1} - \frac{\text{Pe}}{2}\frac{\ell+1}{2\ell+3}a_{\ell+1} \biggr] \text{P}_\ell(\eta) ,
\end{split}
\end{align}
where we used the recursion formula $(\ell + 1)\text{P}_{\ell+1}(\eta) = (2\ell+1) \eta \text{P}_\ell(\eta) - \ell
\text{P}_{\ell -1}(\eta)$. Using the orthogonality relation of the Legendre polynomials,
$\int_{-1}^1\diff \eta\ \text{P}_\ell(\eta) \text{P}_{\ell'}(\eta) = 2\delta_{\ell \ell'}/(2\ell + 1)$, this becomes a tridiagonal matrix equation:
\begin{align} \label{eq:tridiagonal_matrix_equation}
\frac{\kappa \sigma \text{k}_\ell'(\kappa \sigma)}{\text{k}_\ell(\kappa \sigma)} a_\ell 
- \frac{\text{Pe}}{2} \frac{\ell}{2\ell-1}a_{\ell-1} -
\frac{\text{Pe}}{2}\frac{\ell+1}{2\ell+3}a_{\ell+1} = b_\ell . 
\end{align}
The inhomogeneity $b_\ell$ is defined as 
\begin{align}
b_\ell &= \frac{2\ell +1}{2}  ( \text{Pe}  - \img q \sigma  ) \int_{-1}^1\diff\eta\  \eta \text{P}_\ell(\eta)
e^{-\text{Pe}\, \eta  /2}   e^{\img q \sigma \eta} . 
\end{align}
The remaining integral can be be solved using the Rayleigh identity
$ e^{-\img z \cos \vartheta} = \sum_{\ell=0}^\infty (-\img)^\ell (2\ell+1)  \text{j}_\ell(z)
\text{P}_\ell(\cos\vartheta), $
with spherical Bessel function $\text{j}_\ell(\cdot)$ and relying again on the orthogonality relation of the Legendre polynomials:
\begin{align}
b_\ell = -(2\ell +1) ( q \sigma+ \img\text{Pe}) \img^{\ell} \text{j}_\ell'( q
\sigma + \img\text{Pe}/2) ,
\end{align}
where the prime indicates a derivative. 
For the special case of vanishing wavenumber $q = 0$ and the stationary state $\omega = 0$, we recover the tridiagonal matrix
derived in Ref.~\cite{Squires:PoF_17:2005}.

For the forward scattering amplitude~[Eq.~\eqref{eq:forward_scattering_amplitude}] we calculate the integral over the
spherical surface with radius $r = \sigma$:
\begin{align}
\begin{split}
  \frac{1}{\sqrt{V}}\int \diff S_{\vec{r}} \, \img D_r \vec{q} \cdot \hat{\vec{r}} e^{-\img \vec{q}\cdot \vec{r} }
\bigl[\hat{\psi_{q}}(\vec{r},\omega) - \hat{\psi}_{q}^0(\vec{r},\omega)\bigr] 
&= \frac{ 2\pi \img D_r \sigma^2 q /V}{-\img \omega + D_r q^2 + \img \mu F q} \times \\
&\phantom{=}\ \times \sum_{\ell=0}^\infty a_\ell \int_{-1}^1  \diff \eta \, \eta e^{-\img q \sigma \eta } 
e^{\mu F \sigma \eta /2 D_r}  \text{P}_\ell(\eta), 
\end{split}
\end{align}
The integral on the right-hand side can be evaluated again by inserting the Rayleigh identity. As a result, we find
\begin{align}
\begin{split}
\int_{-1}^1  \diff\eta \, \eta e^{-\img q \sigma \eta } 
e^{\eta\text{Pe}/2}  \text{P}_\ell(\eta) = \frac{\partial}{\partial (\text{Pe}/2)} \int_{-1}^1  \diff \eta
\,  e^{-\img q \sigma \eta } 
e^{\eta\text{Pe}/2}  \text{P}_\ell(\eta) 
= 2 \img  (-\img)^\ell \,   \text{j}_\ell'( q \sigma + \img\text{Pe}/2) ,
\end{split}
\end{align}
Collecting results, one finds 
\begin{align}
G(q,\omega) - G_0(q,\omega) 
=  -\frac{4\pi\sigma^3}{3V} G_0(q,\omega) + \frac{4\pi D_r \sigma^2 q}{V} G_0(q,\omega)^2\sum_{\ell=0}^\infty a_\ell (-\img)^\ell \text{j}_\ell'(q\sigma + \img\text{Pe}/2) .
\end{align}
The first term on the right-hand side merely reflects that the free propagator $G_0(q,\omega) = (-\img \omega + D_r q^2 + \img\mu Fq)^{-1}$ allows for particles starting also inside of the obstacle, thereby
renormalizing the residue of the perturbed propagator. We may safely drop this term. 
Conventionally, the result is expressed in terms of a self-energy $\Sigma(q,\omega)$ via the Dyson equation $G = G_0 +
G_0 \Sigma G$.
To first order in the density, the self-energy is merely proportional to the number density $n = N/V$ of the bath particles, and we
obtain the self-energy as our main result of the analytic calculations 
\begin{align} \label{eq:self_energy}
\Sigma(q, \omega) &= 4\pi n D_r \sigma^2 q \sum_{\ell=0}^\infty a_\ell (-\img)^\ell \text{j}_\ell'(q
\sigma + \img \peRe/2) .
\end{align}
The self-energy encodes the density-induced corrections of all moments of the relative motion along
the force. The leading factor $q$ reflects the particle-conservation law.

Without external driving, $\peRe = 0$, the tridiagonal matrix for the coefficients $a_\ell$
[Eq.~\eqref{eq:tridiagonal_matrix_equation}] becomes diagonal yielding
\begin{align}
a_\ell = -(2\ell +1) q \sigma \img^{\ell} \text{j}_\ell'( q \sigma )\frac{\text{k}_\ell(\kappa \sigma)}{\kappa \sigma \text{k}_\ell'(\kappa \sigma)} .
\end{align}
Thus, we recover the known result for the self-energy for equilibrium~\cite{Hanna:PhysA_111:1982,Hanna:JPAMG_14:1982, Ackerson:JCP_76:1982,
Felderhof:PhysA_122:1983}:
\begin{align} \label{eq:self_energy_eq}
\Sigma(q, \omega) &= -4\pi n D_r \sigma^3 q^2 \sum_{\ell=0}^\infty (2\ell +1) \bigl[\text{j}_\ell'( q \sigma )\bigr]^2\frac{\text{k}_\ell(\kappa \sigma)}{\kappa \sigma \text{k}_\ell'(\kappa \sigma)}.
\end{align}

In the case of driving, only terms of order $\ell = \mathcal{O}(\text{Pe})$ significantly contribute for small wave
numbers and the matrix~[Eq.~\eqref{eq:tridiagonal_matrix_equation}] may be safely truncated for numerical evaluation.
The matrix inversion does not generate nonanalytic behavior and one infers that the coefficients $a_\ell \equiv
a_\ell(q\sigma, \kappa\sigma, \peRe)$ should be analytic functions in the arguments. In particular, the matrix is suited for
a perturbative approach for small P\'eclet numbers, in particular, the linear response results can be derived.

The matrix in Eq.~\eqref{eq:tridiagonal_matrix_equation} determines the coefficients $a_\ell$ and thereby the complete
solution of the self-energy~[Eq.~\eqref{eq:self_energy}]. Up to matrix inversion and a temporal Fourier back-transform, which have to be implemented
numerically, we have elaborated a complete time-dependent analytic solution for the relative motion of the probe
particle in the presence of bath particles for the dilute case.

\section{Cumulants of the displacement} \label{sec:cumulants_of_displacement}

The intermediate scattering function generates the moments of the displacement along the field by taking derivatives 
with respect to the wavenumber. In this section we elaborate explicitly how the mean displacement, respectively the 
time-dependent nonlinear mobility, and the fluctuations along the field can be calculated from the self-energy of
the relative motion.

The starting point are the intermediate scattering functions for the displacement~[Eqs.~\eqref{eq:int_scat_prod} and
\eqref{eq:int_scat_gaussian}] where we put the wave vector along the field,
$\vec{q}\parallel\vec{F}$, which is chosen as the $z$ direction. The corresponding cumulant generating function is obtained by taking the logarithm:
\begin{align} 
\begin{split}
\label{eq:cumulant_generating_function}
\ln \langle e^{-\img q \Delta z_a(t)} \rangle 
&= \ln \langle e^{-\img q \Delta Z(t)}\rangle + \ln \langle e^{-\img (D_a q/D_r) \Delta z(t)}\rangle \\
&=-\img q (\mu F D_b/D_r) t  - q^2 (D_a D_b/D_r) t + \ln \langle e^{-\img (D_a q/D_r) \Delta z(t)}\rangle.
\end{split}
\end{align}
This formula establishes the connection between the relative motion $z(t)$ and the displacement of the probe
particle $z_a(t)$.

\subsection{Average motion}

First, we discuss the time-dependent nonlinear mobility defined by
\begin{align} \label{eq:time_dependent_nonlinear_mobility}
\mu(t,\text{Pe}) := \frac{1}{F}\frac{\diff}{\diff t} \langle \Delta z_a(t)\rangle,
\end{align}
characterizing the mean motion of the probe particle, upon switching on the force in the $z$
direction. From the cumulant generating function of the displacement $\Delta z_a(t)$
[Eq.~\eqref{eq:cumulant_generating_function}], the mean-displacement
$\langle \Delta z_a(t) \rangle$ is obtained as
\begin{align}
\begin{split} \label{eq:mean_displacement_center_to_lab}
\langle \Delta z_a(t)\rangle 
&= \img \frac{\partial}{\partial q} \ln\langle e^{-\img q \Delta z_a(t)}\rangle\Bigr|_{q=0} 
= \mu F t D_b/D_r + \frac{\img D_a}{D_r}\frac{\partial}{\partial(D_a q/D_r)} \ln \langle
e^{-\img(D_a q/D_r)\Delta z(t)}\rangle\Bigr|_{q=0} \\
&= \mu F t D_b/D_r + \frac{D_a}{D_r} \langle \Delta z(t) \rangle . 
\end{split}
\end{align}
The contribution from the relative motion can be obtained from the intermediate scattering
function~[Eq.~\eqref{eq:intermediate_scattering_function_r}] as the first derivative $\partial/\partial q|_{q=0}$.  
The dynamics of the relative distance $\Delta z(t)$ between probe and bath particle is contained in the propagator 
\begin{align} \label{eq:propagator_g_first_order}
G(q,\omega) = G_0(q,\omega) + G_0(q,\omega)^2 \Sigma(q,\omega) + \mathcal{O}(n)^2, 
\end{align}
with free propagator $G_0(q,\omega) = (-\img\omega + D_r q^2 +\img q \mu F)^{-1}$ and self-energy
$\Sigma(q,\omega)$~[Eq.~\eqref{eq:self_energy}].
Taking the $q$-derivative leads to the appearance of the 
frequency dependent expression $\mathcal{L}\{\langle \Delta z(t)\rangle\}(\omega) :=
\int_0^\infty \diff t\ e^{\img\omega t} \langle \Delta z(t)\rangle$. It is obtained by considering the propagator $G(q,\omega)$~[Eq.~\eqref{eq:propagator_g_first_order}]:
\begin{align}
\begin{split} \label{eq:relative_mean_displacement_freq}
\mathcal{L}\{\langle \Delta z(t) \rangle\}(\omega) &= \img \frac{\partial}{\partial q} G(q,\omega)\Big|_{q=0} 
= \biggl[\img \frac{\partial G_0}{\partial q} + \img G_0^2\frac{\partial\Sigma}{\partial q}\biggr]_{q=0} \\
&= \frac{\mu F}{(-\img \omega)^2} + 2\pi n \sigma^3 \frac{\mu F}{(-\img\omega)^2} 
\sum_{\ell = 0}^\infty a_\ell \frac{\text{i}_\ell'(\text{Pe}/2)}{\text{Pe}/2} ,
\end{split}
\end{align}
where, in the second line, we used the relation $D_r/\sigma = \mu F/\text{Pe}$
and expressed the spherical Bessel functions $\text{j}_\ell(\cdot)$ in terms of $(-\img) \text{i}'_\ell(\text{Pe}/2) = (-\img)^{\ell}\text{j}'_\ell(\img \text{Pe}/2)$ with modified
spherical Bessel function $\text{i}_\ell(\cdot)$. The coefficients $a_\ell = a_\ell(q\sigma,\kappa\sigma,\text{Pe})$ are to be evaluated for vanishing
wavenumber $q = 0$. In the above equation, the first term corresponds to the free motion and the second
encodes the density-induced response.
Then, the mean displacement of the probe
particle along the field, $\langle \Delta z_a(t)\rangle$, is determined via
Eq.~\eqref{eq:mean_displacement_center_to_lab} leading to
\begin{align}
\begin{split}
\mathcal{L}\{\langle \Delta z_a(t)\rangle\}(\omega) 
&= \frac{D_b}{D_r}\frac{\mu F}{(-\img\omega)^2} + \frac{D_a}{D_r}\mathcal{L}\{\langle \Delta z(t)\rangle\}(\omega) \\
&= \frac{\mu F}{(-\img\omega)^2} + 2\pi n \sigma^3 \frac{\mu F}{(-\img\omega)^2} \frac{D_a}{D_r} \sum_{\ell = 0}^\infty
a_\ell \frac{\text{i}_\ell'(\text{Pe}/2)}{\text{Pe}/2}.  
\end{split}
\end{align}

\begin{figure}
\includegraphics[width=0.65\linewidth]{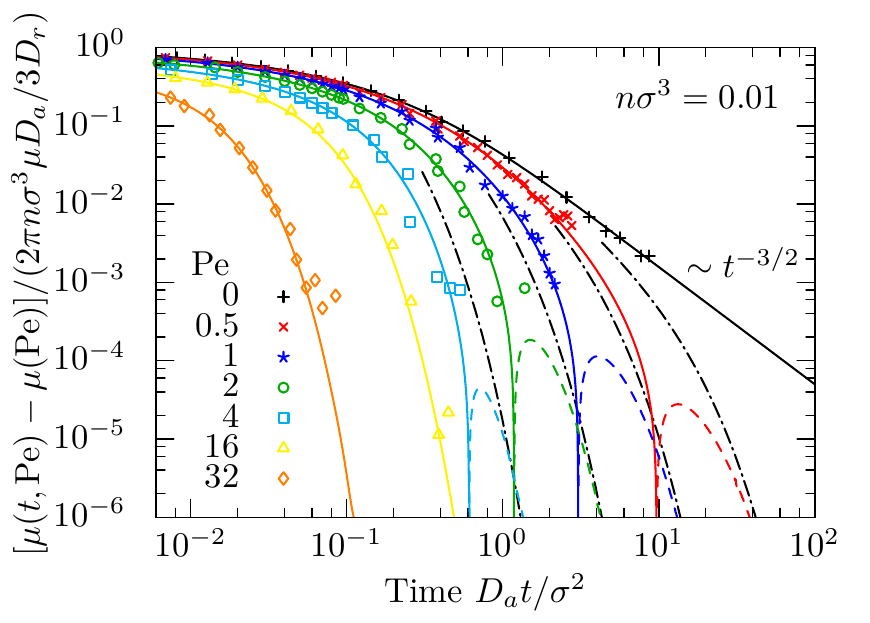}
\caption{\label{fig:time_dependent_mobility}
Time-dependent approach of the mobility $\mu(t,\text{Pe})$ to its stationary-state for different P{\'e}clet numbers,
$\text{Pe}$, and the linear response, $\text{Pe} = 0$.  Solid lines represent the analytic solution and symbols
correspond to Brownian-dynamics simulations for equal sized colloids ($D_b = D_a$). Dashed lines indicate negative
values. The black dashed-dotted lines show the long-time behavior ${\sim-t^{-3/2}\exp(-\text{Pe}^2 D_a t/2\sigma^2)}$ for
the approach to the stationary state.
}
\end{figure}

By a one-sided Fourier transform of the time-dependent nonlinear mobility~[Eq.~\eqref{eq:time_dependent_nonlinear_mobility}],
we obtain the nonlinear frequency-dependent mobility $\hat{\mu}(\omega,\text{Pe}) := -\img\omega\mathcal{L}\{\langle
\Delta z_a(t)\rangle\}(\omega)/F$ with 
\begin{align}
\begin{split} \label{eq:frequency_dependent_mobility}
-\img\omega\hat{\mu}(\omega, \text{Pe})/\mu 
= 1 + 2\pi n \sigma^3\frac{D_a}{D_r} \sum_{\ell = 0}^\infty a_\ell \frac{\text{i}_\ell'(\text{Pe}/2)}{\text{Pe}/2} 
\end{split}
\end{align}

For the time-dependent case, we perform a numerical Fourier inversion of Eq.~\eqref{eq:frequency_dependent_mobility} and
compare to event-driven Brownian dynamics simulations at low density~[Fig.~\ref{fig:time_dependent_mobility}].  Initially, the probe
particle experiences only the drag of the pure solvent, $\mu(t\to 0,\text{Pe}) = \mu$. Then the dynamics slows down due
to the interaction with the bath particles and approaches its stationary mobility $\mu(\text{Pe}) :=
\mu(t\to\infty,\text{Pe})$. Its dependence on the P\'eclet number is discussed in Sec.~\ref{sec:long_time_behavior} below.

In linear response, $\text{Pe} = 0$, we recover the equilibrium solution
[Eq.~\eqref{eq:mobility_equilibrium}] as anticipated by the fluctuation-dissipation theorem. 
In particular, for the equilibrium case, the approach to the stationary state is completely monotone and the time
dependence can be worked out explicitly by Laplace inversion of
Eq.~\eqref{eq:mobility_equilibrium}~\cite{Hanna:JPAMG_14:1982}, leading to
\begin{align}
\mu(t)/\mu = 1 - \frac{2\pi n \sigma^3}{3}\frac{D_a}{D_r}\bigl\{1 - \cos(2t/\tau)[1 - 2\text{C}(\sqrt{4 t/\pi\tau})] -
\sin(2t/\tau)[1-2\text{S}(\sqrt{4 t/\pi\tau})]\bigr\},
\end{align}
with Fresnel integrals $\text{S}(x)=\int_0^x\diff u \sin(\pi u^2/2)$ and $\text{C}(x)=\int_0^x\diff u \cos(\pi
u^2/2)$~\cite{Olver:2010:NHMF,NIST:DLMF}. Inserting the asymptotic expansions for the Fresnel integrals for large
arguments, we recover the algebraic approach of $\sim t^{-3/2}$ to the stationary mobility which is due to repeated
collisions of the probe particle with the same bath particle:
\begin{align}\label{eq:mobility_equilibrium}
\mu(t)/\mu = 1 - \frac{2\pi n \sigma^3}{3}\frac{D_a}{D_r}\Bigl\{1 -
\frac{1}{4\sqrt{\pi}}(t/\tau)^{-3/2} +
\mathcal{O}(t^{-7/2})\Bigr\}, \quad t\to\infty.
\end{align}
The persistent memory in the system also emerges as a long-time tail of the form $\sim t^{-5/2}$ in the
velocity-autocorrelation function, $k_\text{B} T \diff \mu(t) /\diff t$.

For finite driving, expansion of Eq.~\eqref{eq:frequency_dependent_mobility} yields 
\begin{align}
 -\img\omega\hat{\mu}(\omega, \text{Pe})/\mu 
=& 1 - 2\pi n \sigma^3\frac{D_a}{D_r} \Big[ 1 - \frac{\text{Pe}^2}{120}  + \frac{\text{Pe}^2}{8} \kappa \sigma - \frac{1}{2} (\kappa \sigma)^2 + \frac{1}{2} (\kappa \sigma)^3 + {\cal O}(\text{Pe}^4, \kappa^2 \text{Pe}^2, \kappa^4)\Big] 
\end{align}
which is analytic in the complex wavenumber $\kappa$ and the P{\'e}clet number. Moreover only even terms in $\text{Pe}$ appear. The nonanalytic dependence on the frequency and the P{\'e}clet number arises via the square root $\kappa \sigma = \sqrt{-\img \omega \tau + (\text{Pe}/2)^2}$. For vanishing forces $\kappa \sigma = \sqrt{-\img \omega \tau}$ and the term ${\cal O}(\kappa)^3$ generates the long-time tail in Eq.~\eqref{eq:mobility_equilibrium}. For small but finite forces, the singularity is shifted in the complex plane and this term yields  the long-time behavior in the time domain  
\begin{align}\label{eq:mu_small_Pe_long_time}
\frac{\mu(t,\text{Pe}) - \mu(\text{Pe})}{2\pi n \sigma^3 D_a/3D_r} = \frac{1}{4\sqrt{\pi}} (t/\tau)^{-3/2}
\exp(-\text{Pe}^2 t/4\tau), \quad \text{Pe}\to 0 
\quad t\to\infty,
\end{align}
Therefore  the initial decay
becomes more rapid and the long-time tail is followed only for small forces up to some driving-dependent crossover time
$\tau_F := \tau/\text{Pe}^2$, where the tail is decorated by a decaying exponential. Yet, there is a second non-analytic contribution at finite driving ${\cal O}(\text{Pe}^2 \kappa)$ (see also Ref.~\cite{Leitmann:JPhysA:2018} for the corresponding lattice case) which yields in the temporal domain
\begin{align}\label{eq:mu_long_time_small_Pe}
\frac{\mu(t,\text{Pe}) - \mu(\text{Pe})}{2\pi n \sigma^3 D_a/3D_r} = -\frac{1}{8\sqrt{\pi}}  \, \text{Pe}^2 (t/\tau)^{-1/2}
\exp(-\text{Pe}^2 t/4\tau), \quad    t\to\infty ,\quad  \text{Pe}\to 0
\end{align} 
Comparing  expressions in Eqs.~\eqref{eq:mu_small_Pe_long_time},\eqref{eq:mu_long_time_small_Pe} reveals that for times $t\gtrsim \tau_F =  \tau /\text{Pe}^2$ the latter decays more slowly and is the relevant one. Therefore, the divergent time scale $\tau_F$ in the problem separates two different regimes. For $t\lesssim \tau_F$ the linear response prediction remains qualitatively correct, while for $t\gtrsim \tau_F$ the non-equilibirium driving dominates. The second regime also explains the counterintuitive sign change in Fig.~\ref{fig:time_dependent_mobility} such that in the terminal regime the velocity of the probe particle speeds up
again for small and intermediate P{\'e}clet numbers. For large P{\'e}clet number the approach to the stationary state
becomes monotonic again.

The simulations quantitatively confirm the theory for a low density of bath particles (see
Appendix~\ref{appendix:computer_simulations} for simulation details). Note that the simulations were performed at a
finite density of $n\sigma^3 = 0.01$ [Fig.~\ref{fig:time_dependent_mobility}]. Nevertheless, the agreement for the
approach to the stationary behavior of the mobility extends to relative order $10^{-3}$. In particular, for equilibrium,
we observe the persistent power-law correlations due to repeated interactions over roughly one decade in time.

\subsection{Fluctuations along the force} 

The next interesting quantity for the motion of the probe particle are the time-dependent fluctuations around the drift
motion,
\begin{align}
\text{Var}_z(t) := \langle \Delta z_a(t)^2\rangle - \langle \Delta z_a(t)\rangle^2, 
\end{align}
which is the second cumulant of the fluctuating probe displacement $\Delta z_a(t)$. From the cumulant generating
function~[Eq.~\eqref{eq:cumulant_generating_function}], it is obtained via
\begin{align}
\begin{split} \label{eq:varz_transform_center_to_probe}
\text{Var}_z(t) 
&= -\frac{\partial^2}{\partial q^2} \ln \langle e^{-\img q \Delta z_a(t)}\rangle 
= 2(D_a D_b/D_r)t - \frac{D_a^2}{D_r^2}\frac{\partial^2}{\partial (D_a q/D_r)^2} \ln \langle e^{-\img (D_a q/D_r) \Delta
z(t)}\rangle \\
&= 2 (D_a D_b/D_r) t + \frac{D_a^2}{D_r^2} \bigl[\langle \Delta z(t)^2\rangle - \langle \Delta z(t) \rangle^2\bigr],
\end{split}
\end{align}
where, in the second line, we already inserted the explicit expression for the second cumulant of the relative
displacement $\Delta z(t)$.
First, we calculate the second moment along the field in the frequency-domain:
\begin{align}
\begin{split}
\mathcal{L}\{\langle \Delta z(t)^2\rangle\}(\omega) &= -\frac{\partial^2}{\partial q^2} G(q,\omega)\biggr|_{q=0} = \biggl[-\frac{\partial^2 G_0}{\partial q^2} - 4 G_0 \frac{\partial
G_0}{\partial q}\frac{\partial \Sigma}{\partial q} - G_0^2\frac{\partial^2\Sigma}{\partial q^2}\biggr]_{q=0}
\\
&= \frac{2 D_r}{(-\img\omega)^2} + \frac{2(\mu F)^2}{(-\img\omega)^3} 
+ 2\pi n \sigma^3 \frac{4(\mu F)^2}{(-\img\omega)^3} \Delta\hat{\mu}(\omega,\text{Pe}) + 
 2\pi n \sigma^3 \frac{4D_r}{(-\img\omega)^2} \Delta \hat{R}(\omega, \text{Pe}). 
\end{split}
\end{align}
where we introduce auxiliary functions
\begin{align}
\Delta \hat{\mu}(\omega,\text{Pe}) &:= \sum_{\ell = 0}^\infty a_\ell \frac{\text{i}_\ell'(\text{Pe}/2)}{\text{Pe}/2}
\label{eq:response_func_mobility} , \\
\Delta \hat{R}(\omega, \text{Pe}) &:= \sum_{\ell = 0}^\infty \Bigl[\frac{\img}{\sigma} \frac{\partial a_\ell}{\partial
q} \text{i}_\ell'(\text{Pe}/2) + a_\ell\text{i}_\ell''(\text{Pe}/2)\Bigr] \label{eq:response_func_msd} .
\end{align}
In particular, with the auxiliary function for the mobility, $\Delta \hat{\mu}(\omega,\text{Pe})$, the
frequency-dependent nonlinear mobility~[Eq.~\eqref{eq:frequency_dependent_mobility}] can be written as
\begin{align}
-\img\omega\hat{\mu}(\omega, \text{Pe})/\mu = 1 + 2\pi n \sigma^3\frac{D_a}{D_r} \Delta \hat{\mu}(\omega,\text{Pe}) .
\end{align}
In the time-domain, the mean-square displacement then follows as
\begin{align}
\begin{split}
\langle \Delta z(t)^2\rangle &= 2 D_r t + (\mu F)^2 t^2 
+ 8 \pi n \sigma^3 D_r \int_0^t\mathrm{d} t'\ \mathcal{L}^{-1}\biggl\{\frac{\Delta \hat{R}(\omega, \text{Pe})}{(-\img\omega)} +
\frac{(\mu F)^2}{D_r} \frac{\Delta \hat{\mu}(\omega, \text{Pe})}{(-\img\omega)^2}\biggr\}(t') ,
\end{split}
\end{align}
where we introduced the inverse Laplace transform $\mathcal{L}^{-1}\{\cdot\}(t)$. 
To calculate the square of the mean displacement $\langle \Delta z(t)\rangle$, we first express the frequency-dependent
mean displacement given in Eq.~\eqref{eq:relative_mean_displacement_freq} in the time-domain leading to
\begin{align}
\langle \Delta z(t) \rangle = \mu F t + 2\pi n \sigma^3 \mu F  \int_0^t \mathrm{d} t'\
\mathcal{L}^{-1}\biggl\{\frac{\Delta \hat{\mu}(\omega, \text{Pe})}{(-\img\omega)}\biggr\}(t').
\end{align}
Then, the square of the mean-displacement to first order in the density is given by
\begin{align}
\langle \Delta z(t) \rangle^2 = (\mu F)^2 t^2 + 4 \pi n\sigma^3(\mu F)^2 t \int_0^t \mathrm{d}
t'\ \mathcal{L}^{-1}\biggl\{\frac{\Delta\hat{\mu}(\omega, \text{Pe})}{(-\img\omega)}\biggr\}(t') + \mathcal{O}(n^2).
\end{align}
Collecting results, the time-dependent diffusion coefficient along the field, 
$D_z(t,\text{Pe}) := (1/2)\mathrm{d} \text{Var}_z(t)/ \mathrm{d} t$, 
can be written as 
\begin{align}
\begin{split} \label{eq:time_dependent_diff_coeff}
D_z(t,\text{Pe}) 
&= D_a + 2\pi n \sigma^3 \frac{D_a^2}{D_r} \mathcal{L}^{-1}\biggl\{\frac{2\Delta \hat{R}(\omega, \text{Pe})}{(-\img\omega)} + 
 \frac{\text{Pe}^2}{(-\img\omega)} \frac{\partial \Delta \hat{\mu}(\omega, \text{Pe})}{\partial
(-\img\omega\tau)}\biggr\}(t),
\end{split}
\end{align}
where we used the relation $t \mathcal{L}^{-1}\{(\cdot)\}(t) = \mathcal{L}^{-1}\{\partial (\cdot)/ \partial(-\img\omega)\}(t)$ of the
Laplace transform.

\begin{figure}
\includegraphics[width=0.65\linewidth]{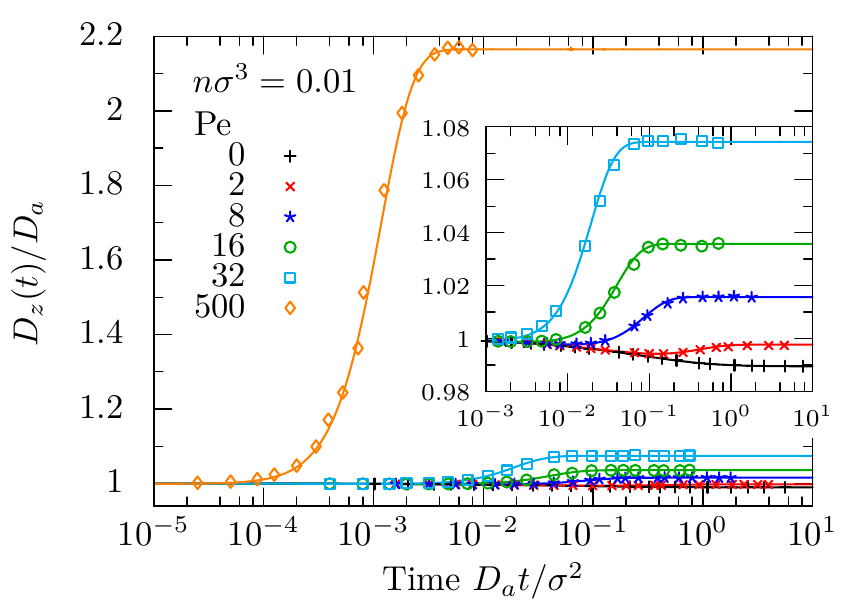}
\caption{\label{fig:diff_coeffz}
Time-dependent diffusion coefficient $D_z(t)$ for different strength of the driving, $\text{Pe} = \mu F \sigma/D_r$.
Lines correspond to the analytic solution and symbols represent Brownian-dynamics simulation for equal sized colloids
($D_b = D_a$). Inset: Zoom of the same quantity for the smaller P{\'e}clet numbers.}
\end{figure}

For small times the probe particle does not interact with the bath
particles and the diffusion coefficient is given by the bare one $D_z(t\to 0,\text{Pe}) = D_a$. Only in the equilibrium
case, the diffusion coefficient decreases monotonically~[see Fig.~\ref{fig:diff_coeffz}] to its stationary-state value
$D_z^\text{eq}/D = 1 - (D_a/D_r) 2\pi n \sigma^3/3$. 

For any finite driving, a minimum of least diffusivity emerges at intermediate times such that the growth of the
fluctuations speeds up again until the stationary diffusion coefficient is reached~[Fig.~\ref{fig:diff_coeffz} (inset)].
With increasing driving the time of least diffusivity becomes smaller and smaller. The time-dependent growth becomes
arbitrarily large as the P{\'e}clet number is increased, even at small densities. A growing time-dependent diffusion
coefficient is a fingerprint of transient superdiffusive behavior. 

\begin{figure}
\includegraphics[width=0.65\linewidth]{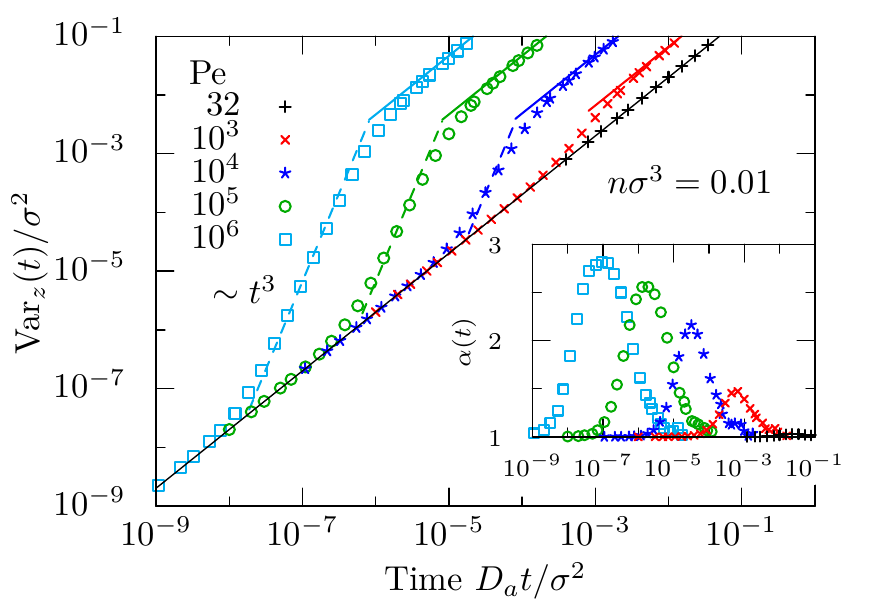}
\caption{\label{fig:varz_colloid}
Time-dependent fluctuations $\text{Var}_z(t) = \langle\Delta z_a(t)^2\rangle - \langle \Delta z_a(t)\rangle^2$ and local
exponent $\alpha(t) = \diff\!\ln[\text{Var}_z(t)]/\diff\!\ln(t)$ (inset) of the
probe particle along the applied force obtained from Brownian dynamcis simulation of equal sized colloids ($D_b = D_a$) for different strength of the driving. The solid lines correspond to the diffusive
asymptote $\text{Var}_z(t) = 2D_z(\text{Pe})t$ and the dashed lines are the asymptotic model
$\text{Var}_z(t) = (D_a/D_r)^2 (\mu F t)^3/3l_*$ with the emperical mean-free path length $l_* \approx 1/n \sigma^2$.
}
\end{figure}

The superdiffusion becomes more and more pronounced upon increasing the force beyond values that are feasible for
numerical implementation of our analytical solution ($\text{Pe} \gtrsim 10^3$). Nevertheless, in simulations the regime
of these large P{\'e}clet numbers can still be accessed~[Fig.~\ref{fig:varz_colloid}]. The window of superdiffusion
opens for stronger driving and the local exponent for the variance 
\begin{align}
\alpha(t) := \frac{\diff\!\ln[\text{Var}_z(t)]}{\diff\!\ln(t)} = \frac{2 D_z(t,\text{Pe}) t}{\text{Var}_z(t)}
\end{align}
approaches a value of $3$ at intermediate times. The long-time behavior is for all forces diffusive with a strongly
enhanced diffusion coefficient $D_z(\text{Pe})/D \simeq \pi n \sigma^3 \text{Pe} [\ln(2)-1/4]/6$, for $\text{Pe}\to\infty$~\cite{Zia:JFM_658:2010}.

Let us rationalize the superdiffusive behavior in terms of an asymptotic model for the limit of strong driving. Therefore we adapt an earlier 
asymptotic model valid for the driven lattice Lorentz model~\cite{Leitmann:PRL_118:2017}.
Here,
the probe particle's motion is dominated by the drift with constant velocity $\mu F$ along the force until it hits a
bath particle for the first time and then slowly slides along its surface.  In this time regime, the motion
becomes essentially one-dimensional such that the probability distribution for the relative motion, $\mathbb{P}(\Delta
z, t)$, consists of freely moving particles with fixed velocity or particles that are transiently blocked by bath particles.
The free path lengths are exponentially distributed, since at
low density the positions of the bath particles are independent, and the probability distribution can be estimated
directly to
\begin{align} \label{eq:distribution_displacements_asymptotic}
\mathbb{P}(\Delta z, t) = \delta(\Delta z - v t)e^{-\Delta z/l_*} + (1/l_*) e^{-\Delta z/l_*}\vartheta(vt - \Delta z),
\end{align}
where $l_* = 1/n\Sigma_*$ denotes a mean-free path length. The first term corresponds to the freely moving probe (with
$v = \mu F$), while the second term accounts for the blocked probe particles. 

From the distribution of the
displacements~[Eq.~\eqref{eq:distribution_displacements_asymptotic}], it is straight forward to calculate the
mean and the mean-square displacement. Then, the growth of the fluctuations for the probe particle are obtained from
Eq.~\eqref{eq:varz_transform_center_to_probe}, where we discarded the diffusive contribution since we are only
interested in the drift motion. As a result, the distribution of the displacements implies a strong initial growth of the variance
with 
\begin{align}
\text{Var}_z(t) = \Bigl(\frac{D_a}{D_r}\Bigr)^2 \frac{(\mu F t)^3}{3l_*},
\end{align}
which is also confirmed by Brownian dynamcis simulation~[Fig.~\ref{fig:varz_colloid}].
Empirically we find, that the relevant
scattering cross section $\Sigma_* \approx \sigma^2$ is smaller than the geometric cross section $\pi\sigma^2$, leading to a mean-free path length of $l_* \approx 1/n \sigma^2$. 
The smaller scattering cross section can be readily interpreted 
 since only head-on collisions effectively stop the directed motion [in the lattice variant no empirical correction was necessary, the scattering cross section coincides with the geometrical one~\cite{Leitmann:PRL_118:2017}].
 Matching the superdiffusion with the
short-time asymptote $2 D_a t$ yields as crossover time $\sim \text{Pe}^{-3/2}$. Similarly the terminal time of
superdiffusion is set by the time the probe particle needs to pass a bath particle $\sigma/\mu F\sim 1/\text{Pe}$
leading by crossover matching to the scaling prediction $D_z(\text{Pe})~\sim n\,\text{Pe}$ for the diffusion coefficient
consistent with Ref.~\cite{Zia:JFM_658:2010}. Thus, the window of superdiffusion expands as $\text{Pe}^{1/2}$ as the
force is increased.

\subsection{Long-time behavior} \label{sec:long_time_behavior}

\begin{figure}
\includegraphics[width=0.65\linewidth]{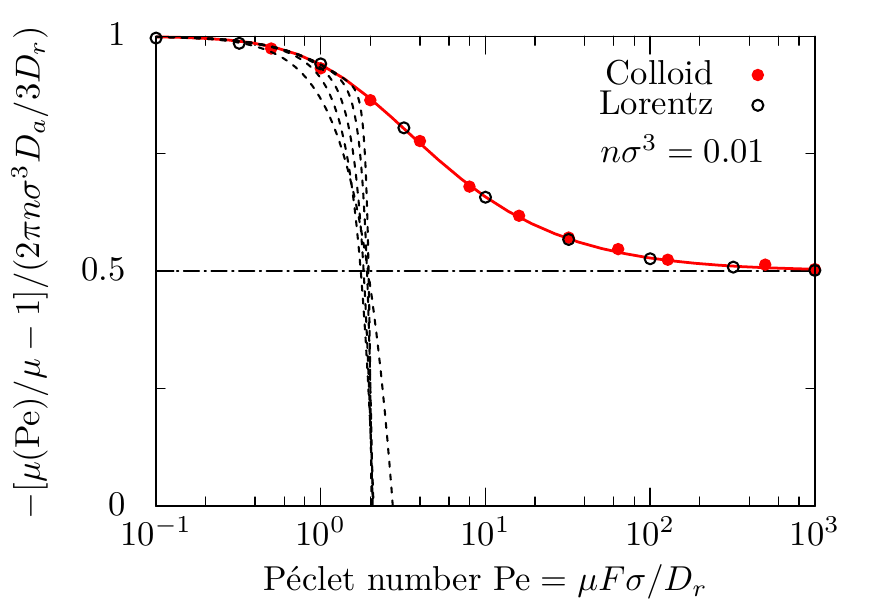}
\caption{\label{fig:stationary_state_mobility}
Density-induced suppression of the stationary mobility
$\mu(\text{Pe})$. Dashed lines indicate the asymptotic expansion for increasing order ($2$, $4$, $6$, $10$, and $20$). 
Lines correspond to the theory and symbols represent Brownian-dynamics simulations of equal sized colloids,
$D_b = D_a$, and Lorentz systems with $D_b = 0$.
}
\end{figure}

Let us specialize our time-dependent solution to the stationary state. 
The stationary mobility $\mu(\text{Pe})$ follows
as special case from Eq.~\eqref{eq:frequency_dependent_mobility} via 
\begin{align} \label{eq:stationary_state_mobility}
\mu(\text{Pe})/\mu = \lim_{\omega\to 0} (-\img\omega) \hat{\mu}(\omega, \text{Pe})/\mu 
= 1 + 2\pi n \sigma^3 \frac{D_a}{D_r} \Delta\mu_0(\text{Pe}) ,
\end{align}
where the coefficient $\Delta\mu_0(\text{Pe})$ is defined via a low-frequency expansions of the auxiliary function 
\begin{align} \label{eq:response_func_mobility_small_freq}
\Delta \hat{\mu}(\omega,\text{Pe}) = \Delta\mu_0(\text{Pe}) + (-\img\omega\tau)\Delta\mu_1(\text{Pe}) + \dotsb,
\quad \omega\to 0.
\end{align}
For small driving and small frequency, we can determine the coefficients $a_\ell$ of the auxiliary function $\Delta
\hat{\mu}(\omega,\text{Pe})$~[Eq.~\eqref{eq:response_func_mobility}] by an inversion of the tridiagonal
matrix equation~[Eq.~\eqref{eq:tridiagonal_matrix_equation}] and by considering the frequency-independent contribution:
\begin{align}
\Delta \mu_0(\text{Pe}) = -\frac{1}{3} + \frac{2}{45}\text{Pe}^2 - \frac{1}{24}|\text{Pe}|^3 +
\frac{128}{4725}\text{Pe}^4 + \mathcal{O}(|\text{Pe}|^5),\quad \text{Pe}\to 0.
\end{align}
Inserting this result into Eq.~\eqref{eq:stationary_state_mobility}, we recover the previously derived asymptotic
expansion~[Eq.~\eqref{eq:series_expansion_squires}]~\cite{Squires:PoF_17:2005}.
In particular, in equilibirium, $\Delta \mu_0(\text{Pe}\to 0) = -1/3$ and for equal-sized colloids with relative
diffusion coefficient $D_r = 2D_a$ and packing fraction $\varphi = \pi n \sigma^3/6$, we
obtain the known result from equilibrium, $\mu(\text{Pe} = 0)/\mu = 1 - 2\varphi$~\cite{Hanna:PhysA_111:1982}.
In the limit of large forces, $\text{Pe}\to\infty$, the analytic solution for the
mobility~[Eq.~\eqref{eq:stationary_state_mobility}] approaches the limit $\mu(\text{Pe}\to\infty)/\mu = 1-(D_a/D_r)\pi n
\sigma^3/3$~[Fig.~\ref{fig:stationary_state_mobility}], derived earlier
in terms of a boundary layer analysis~\cite{Squires:PoF_17:2005}.

Similarly, for the stationary diffusion coefficient $D_z(\text{Pe})$, we
perform the limit of small frequencies in Eq.~\eqref{eq:time_dependent_diff_coeff} and obtain
\begin{align}
D_z(\text{Pe})/D_a = \lim_{\omega\to 0} (-\img\omega) \hat{D}_z(\omega,\text{Pe})/D_a = 1 + 2\pi n \sigma^3
\frac{D_a}{D_r}[2 \Delta R_0(\text{Pe}) + \text{Pe}^2 \Delta \mu_1(\text{Pe})],
\end{align}
where the coefficient $\Delta \mu_1(\text{Pe})$ is defined in Eq.~\eqref{eq:response_func_mobility_small_freq} and the
coefficient $\Delta R_0(\text{Pe}) := \Delta \hat{R}(\omega \to 0,\text{Pe})$
is the small-frequeny limit of the auxiliary function $\Delta \hat{R}(\omega, \text{Pe})$~[Eq.~\eqref{eq:response_func_msd}].
Similar to the coefficient $\Delta \mu_0(\text{Pe})$, we determine the coefficients $a_\ell$ and $\partial
a_\ell/\partial q$ by matrix inversion of the tridiagonal matrix equation~[Eq.~\eqref{eq:tridiagonal_matrix_equation}]
in powers of the the P\'eclet number and the frequency. The coefficients are then obtained as 
\begin{align}
\Delta R_0(\text{Pe}) &= -\frac{1}{6} + \frac{17}{270}\text{Pe}^2 - \frac{11}{144}|\text{Pe}|^3 +
\frac{1679}{28350}\text{Pe}^4 + \mathcal{O}(|\text{Pe}|)^5,\quad \text{Pe}\to 0 ,\\
\Delta\mu_1(\text{Pe}) &= \frac{1}{6} - \frac{1}{6}|\text{Pe}| + \frac{101}{810}\text{Pe}^2 -
\frac{17}{216}|\text{Pe}|^3 + \frac{310571}{6804000}\text{Pe}^4 +
\mathcal{O}(|\text{Pe}|)^5, \quad \text{Pe}\to 0 ,
\end{align}
and the series expansion of the diffusion coefficient parallel to the applied field is calculated to 
\begin{align}
\begin{split} \label{eq:series_diff_coeff}
D_z(\text{Pe})/D_a = 1  - \frac{2\pi n \sigma^3}{3}\frac{D_a}{D_r} \biggl[1 &- \frac{79}{90}\peRe^2 + \frac{23}{24}|\peRe|^3
- \frac{6893}{9450}\peRe^4 + \mathcal{O}(|\peRe|)^5\biggr] .
\end{split}
\end{align}
With our solution, we recover the leading correction $\mathcal{O}(\text{Pe}^2)$ to the equilibrium case, which has been calculated
earlier~\cite{Zia:JFM_658:2010}. Furthermore, our calculation extends this result to arbitrary order and reveals the
emergence of nonanalytic contributions similar to the mobility in the stationary state.

\begin{figure}
\includegraphics[width=0.65\linewidth]{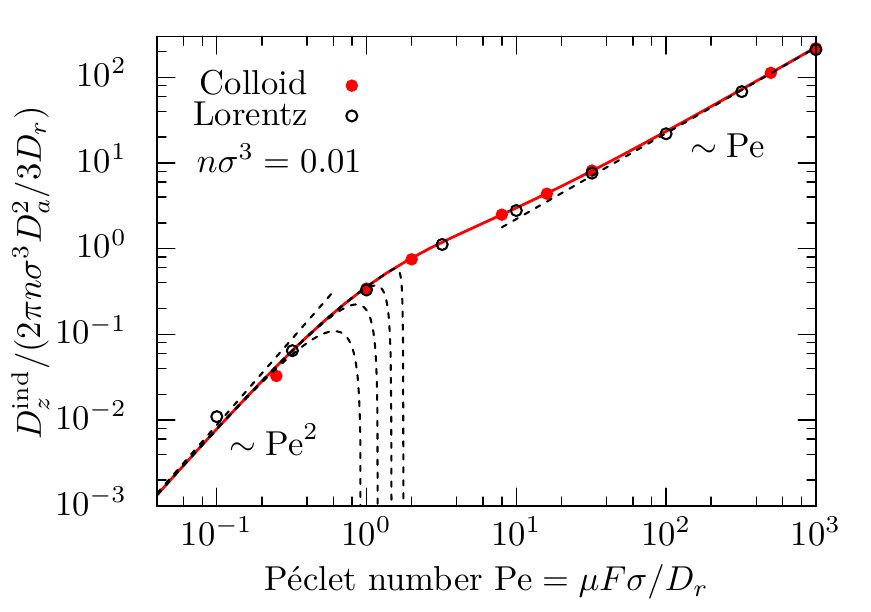}
\caption{\label{fig:stationary_state_diffcoeff}
Stationary-state force-induced diffusion coefficient $D_z^\text{ind} := D_z(\text{Pe}) - D_z(\text{Pe}\to 0)$. Dashed lines represent the
asymptotic behavior for small (increasing order $2$, $3$, $5$, $9$, $31$)  and large driving.
Lines correspond to the theory and symbols represent Brownian-dynamics simulations of equal sized colloids,
$D_b = D_a$, and Lorentz systems $D_b = 0$.
}
\end{figure}

As can be inferred from the stationary mobility~[Fig.~\ref{fig:stationary_state_mobility}] and the stationary diffusion
coefficient~[Fig.~\ref{fig:stationary_state_diffcoeff}], both series expansions
[Eqs.~\eqref{eq:series_expansion_squires} and
\eqref{eq:series_diff_coeff}] break down already at moderate driving. However, the numerical evaluation via the matrix
inversion [Eq.~\eqref{eq:tridiagonal_matrix_equation}] is valid for the full range of P\'eclet numbers which has been  shown first by Khair and Brady~\cite{Khair:JFM_557:2006} for the mobility.
The limiting  value of the stationary mobility as well as the asymptotic behavior $\mathcal{O}(\text{Pe})$ of the stationary diffusion
coefficient for large P\'eclet numbers~[Fig.~\ref{fig:stationary_state_diffcoeff}]~\cite{Zia:JFM_658:2010} is nicely corroborated in the Brownian dynamics simulations.

\subsection{Time-dependent  pair-distribution function}

Our solution provides also the time-dependent  pair-distribution function $g(\vec{r},t)$ for the relative distance $\vec{r}$ of the tracer and the bath particles  by integrating the conditional probability distribution $\psi(\vec{r},t | \vec{r}')$ over all $\vec{r}'$. 
In the frequency domain,  $\hat{g}(\vec{r},\omega) = \hat{\psi}_{q=0}(\vec{r},\omega) \sqrt{V}$  and with Eq.~\eqref{eq:solution_Helmholtz_equation}, we find for $|\vec{r}| > \sigma$
\begin{equation}
-\img \omega  \hat{g}(\vec{r},\omega) = 1+  e^{\mu\vec{F}\cdot \vec{r}/2D_r}   \sum_{\ell=0}^\infty a_\ell \frac{\text{k}_\ell(\kappa r)}{\text{k}_\ell(\kappa \sigma)} \text{P}_\ell(\cos \vartheta),
\end{equation}
where the coefficients $a_\ell \equiv a_\ell (q\sigma, \kappa \sigma, \text{Pe})$ are to be evaluated at zero wavenumber $q=0$. In particular, in equilibrium all coefficients vanish $a_\ell = 0$ and the pair-distribution function reduces to the stationary step function.

\begin{figure}
 \includegraphics[width=0.65\linewidth]{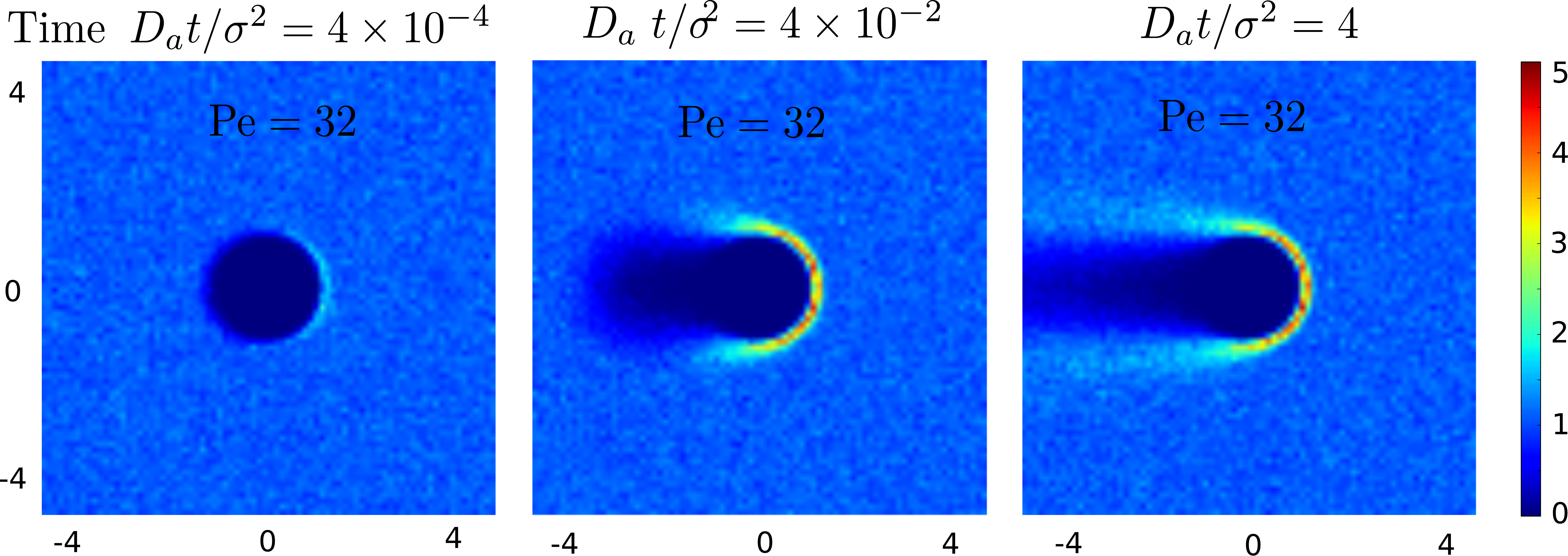} \\
\includegraphics[width=0.22\linewidth]{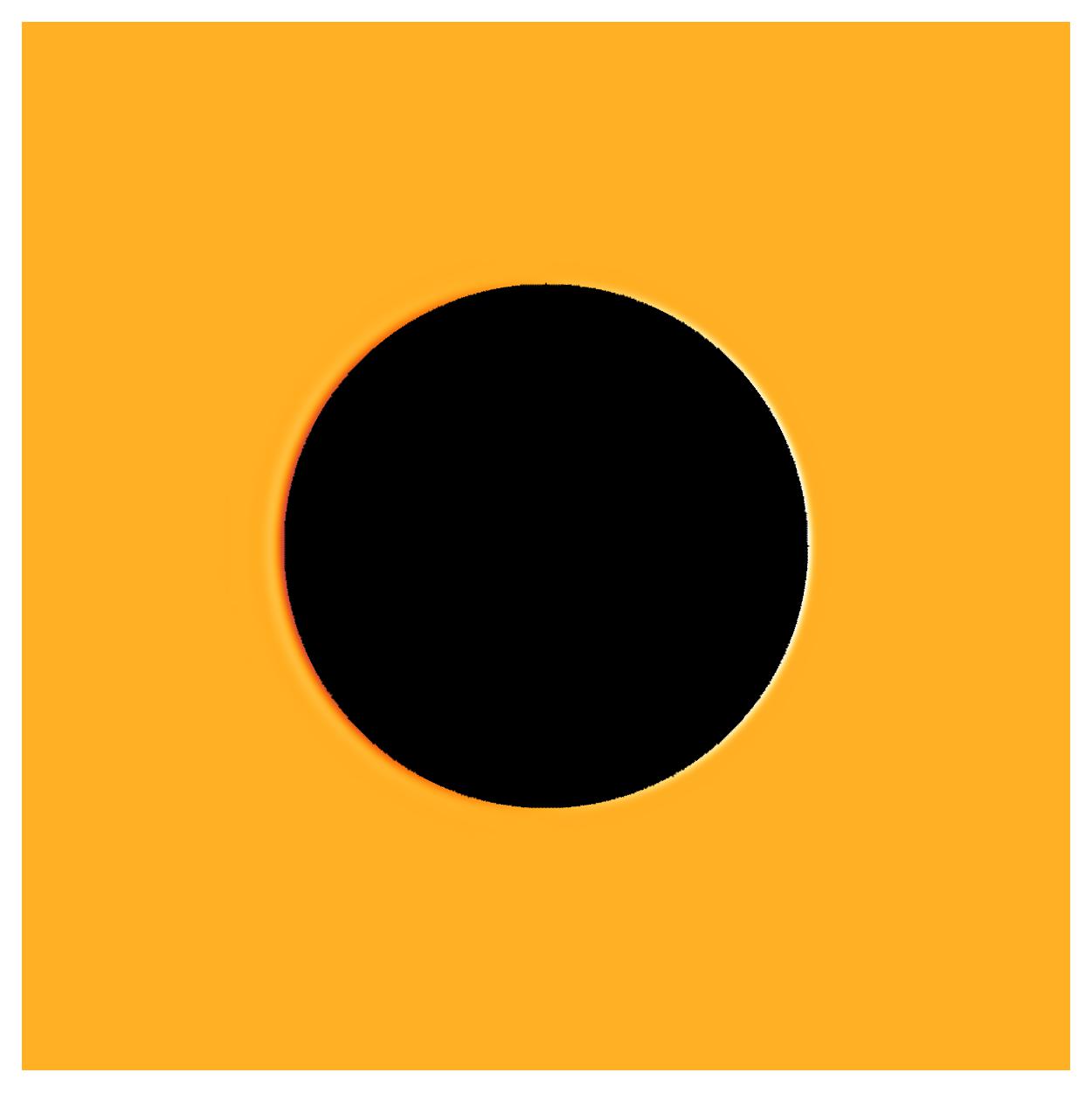}\includegraphics[width=0.22\linewidth]{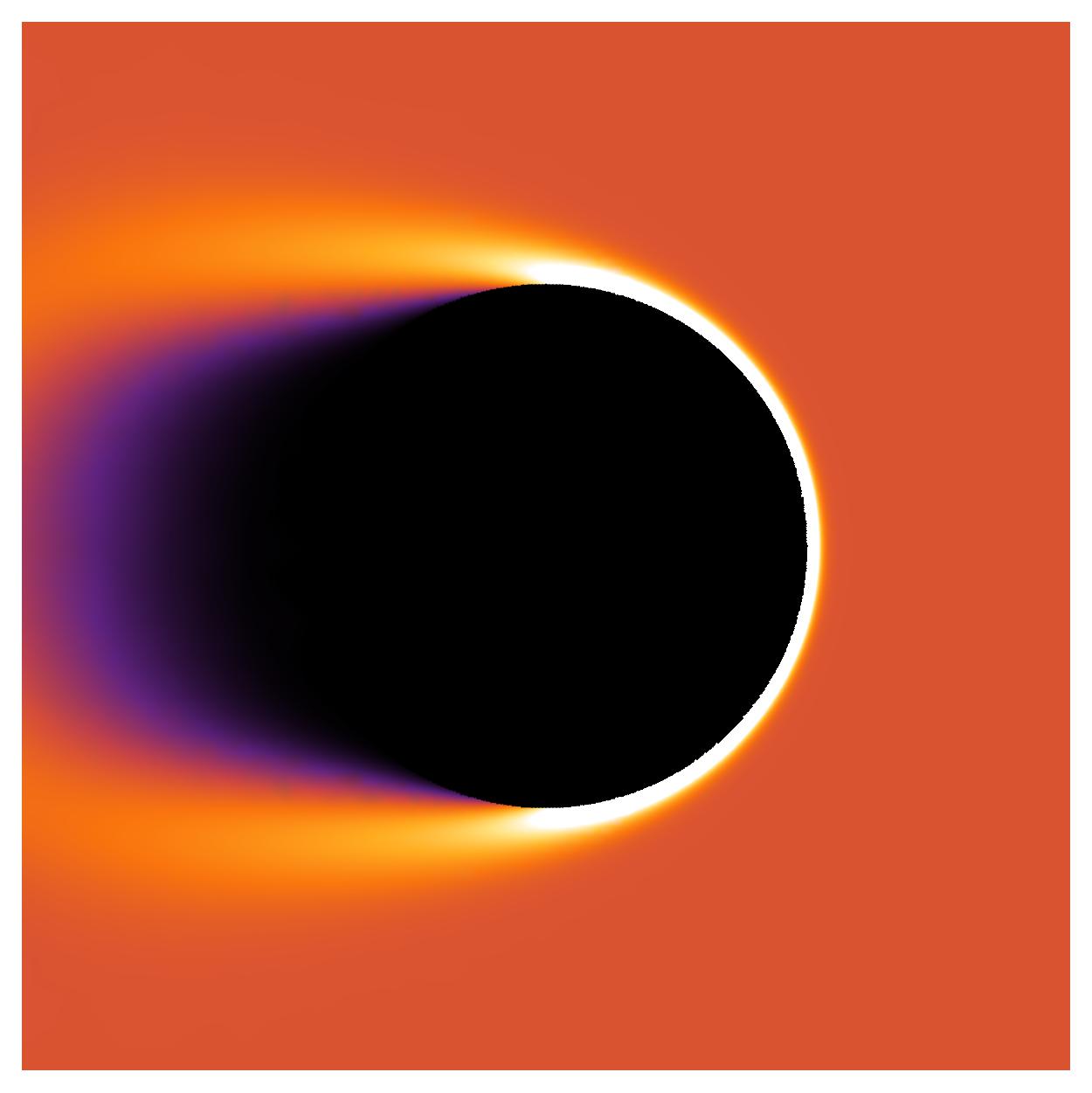}
\includegraphics[width=0.22\linewidth]{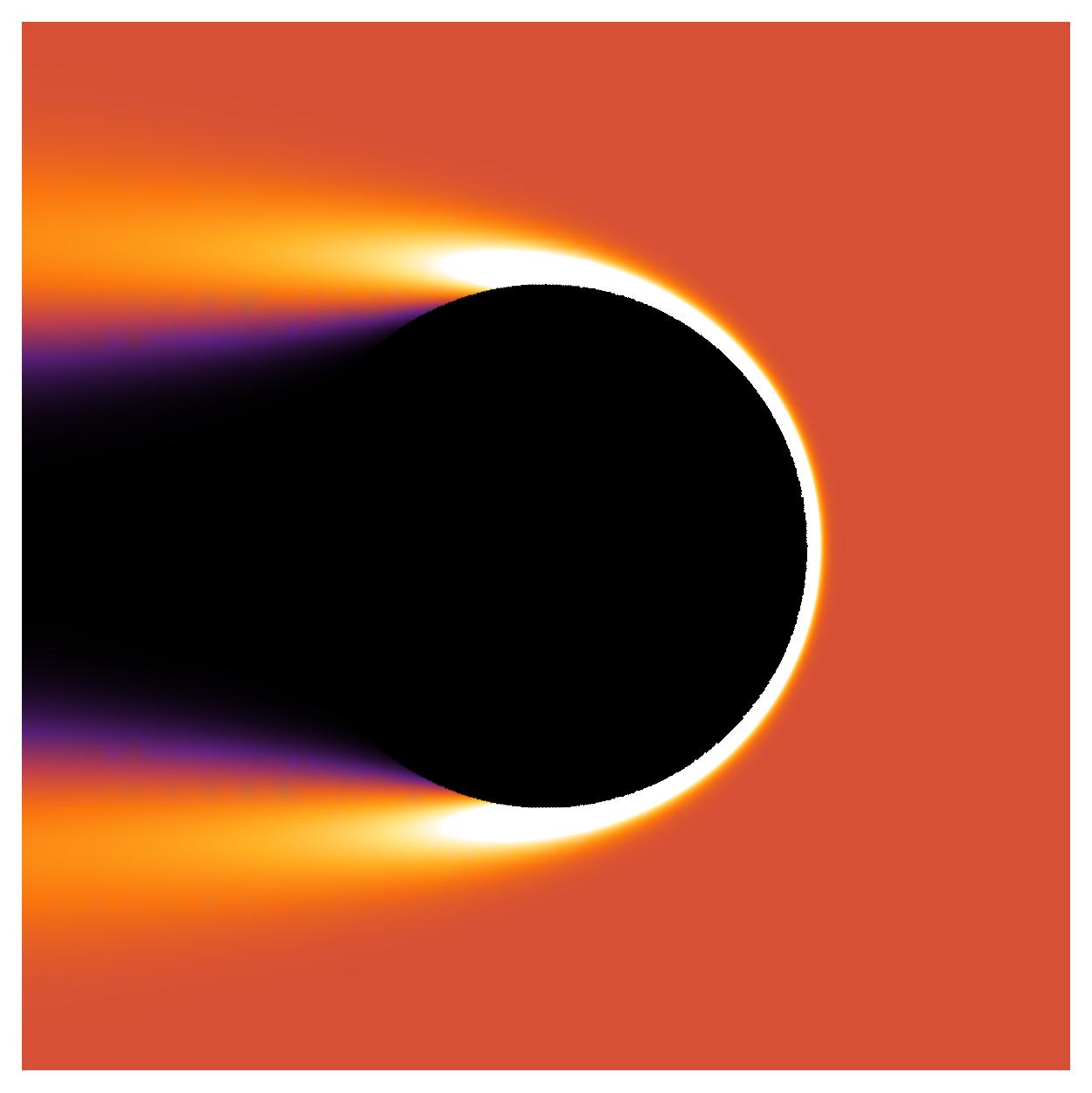}
 \\
\caption{\label{fig:time_dependent_probability_density}
Top: Simulation results for the time-dependent pair-distribution function $g(\vec{r},t)$ for P{\'e}clet number $\text{Pe}=32$ and density $n\sigma^3 = 0.01$. As time progresses $t \cdot D_a/\sigma^2 = 4 \cdot 10^{-4}, 4 \cdot 10^{-2}, 4$ (left to right) the tracer piles up probability in front and leaves a trail of depleted probability. 
Bottom: Analytic result for the real part of the  pair-distribution function in the frequency domain $\Real[-\img \omega \hat{g}(\vec{r},\omega)]$ at the same driving $\text{Pe}=32$ for frequencies $\omega \cdot \sigma^2/2\pi D_a = 1/( 4\cdot 10^{-4}), 1/(4\cdot 10^{-2}), 1/4$ (left to right).  }
\end{figure}

Simulation results for the time-evolution of the pair-distribution function are displayed in Fig.~\ref{fig:time_dependent_probability_density}. Shortly after switching on the force, the pair-distribution function is still almost spherically symmetric, there has been no time to propagate the information of the strong perturbing force to the surroundings. Quickly the tracer's motion is obstructed by the bath particles such that probability accumulates in front of the tracer particle. For longer times, a trailing wake evolves where probability is depleted behind the tracer particle.
The numerical results for the pair-distribution function in the frequency domain shown in Fig.~\ref{fig:time_dependent_probability_density}  corroborate this picture. For small frequencies we recover the stationary distribution calculated in Refs.~\cite{Squires:PoF_17:2005,Khair:JFM_557:2006}.  For high P{\'e}clet numbers the stationary two-particle distribution function becomes strongly asymmetric as has been shown first  Ref.~\cite{Squires:PoF_17:2005} and 
more systematically in Ref.~\cite{Khair:JFM_557:2006}. Probability piles up in a narrow boundary layer in front of the pulled particle particle and leaves behind a wake extending to larger and larger distances ${\cal O}(\text{Pe})$ as the driving force is increased.

\section{Summary and conclusion} \label{sec:summary_and_conclusion}

We have derived an analytic solution for the full time-dependent response of a probe particle driven out of equilibrium by
a step force in first order of the density of bath particles (discarding inertial and hydrodynamic effects). 
The response is completely encoded in the self-energy from
which, in principle, all moments along the force can be generated. In comparison to the known self-energy in equilibrium, the
force results in a complex shift of the frequency $-\img\omega\tau \mapsto -\img\omega\tau + (\text{Pe}/2)^2 =
\kappa^2\sigma^2$ encoded in the complex wavenumber $\kappa\sigma$, an explicit regular variation of the expansion
coefficients $a_\ell$ via the boundary
condition, and a shift in the wavenumber $q\sigma \mapsto q\sigma + \img\,\text{Pe}/2$. 
The shift in the frequency reveals that the nonanalytic frequency behavior in equilibrium, as manifested in the long-time
tails, and the nonanalytic dependence on the driving, are merely two sides of the same coin. In particular, this explains
why at any finite driving the approach of the nonlinear mobility to its stationary value becomes exponentially fast.
Furthermore, this reveals that for finite times, all response functions are analytic functions in 
the driving, however, this does not hold for infinite times since the limits do not commute. 

The emergence of non-analytic behavior and a divergent time scale for $\text{Pe}\to 0$  calls for an  explanation in terms of physics. The stationary Smoluchowski equation is non-uniform at small P{\'e}clet number [the same physics has been discussed for the advection-diffusion equation in the seminal contribution by Acrivos and Taylor~\cite{Acrivos:1962}]: it displays an outer region at distances $r\gg \sigma /\text{Pe}$ where advection dominates, and an inner one where diffusion is the dominant contribution.  For the transport coefficient already to order ${\cal O}(\text{Pe}^2)$
 both regions need to be calculated and matching the solutions  makes the non-analytic contributions evident~\cite{Khair:JFM_557:2006}. For non-zero frequencies or finite times the advected Smoluchowski equation is regular and 
solutions decay exponentially fast on the 'Skin penetration' depth $\sqrt{ D_r/\omega} \sim \sqrt{D_r t}$. Correspondingly, if the Skin penetration depth is smaller than the inner region, i.e. for times $ \sqrt{ D_r t}, \lesssim \sigma/\text{Pe}$, i.e. $t \lesssim \tau_F = \tau/ \text{Pe}^2$,   one can safely ignore the presence of the outer region and the system behaves as in linear response. In contrast, for large enough times, the Skin penetration depth covers the outer region and the nonlinearities becomes important. The interplay of the divergent boundary layer and the Skin penetration depth is the origin of non-commuting limits. Mathematically related, but not quite identical, is the problem of the time-dependent motion at small but finite Reynolds number~\cite{Lovalenti_JFM:1993}.

The full solution provides the first direct access to the time-dependent fluctuations along the force. Here
we explicitly characterized the transient superdiffusion which connects the short-time bare diffusion and the long-time
enhanced diffusion~\cite{Zia:JFM_658:2010}. The emergence of superdiffusion has been rationalized by considering the
distribution of the free path lengths.

Let us comment also on hydrodynamic interactions. Progress has been made for certain limiting cases: In
equilibrium, the stationary diffusion coefficients have been estimated by including instantaneous hydrodynamic
interactions at the level of the Oseen tensor, as well as including near-field corrections. Results for the
density-induced suppression depend somewhat on the approximation of the hydrodynamic interactions but are close to the
diffusion coefficient neglecting hydrodynamics~\cite{Hanna:PhysA_111:1982,Batchelor:JFM_74:1976,Rallison:JFM_167:1986,Hoh:JFM_795:2016}. For the driven case, hydrodynamic interactions have been accounted for in the stationary state both for the structure deformation as well as the mobility for all P{\'e}clet numbers~\cite{Khair:JoR_49:2005}. 
The force-induced corrections
to the stationary diffusion coefficient have been elaborated only recently, and it has been shown that hydrodynamic
interactions do not lead to qualitatively new behavior, although the numerical values change. In particular, the low-force
corrections are still $\mathcal{O}(\text{Pe}^2)$, and in the regime of strong forces they still scale as $\mathcal{O}(\text{Pe})$, however
with a slow convergence to the asymptotic result~\cite{Hoh:JFM_795:2016}.

A second effect due to hydrodynamics arises due to frequency-dependent hydrodynamic interactions implying hydrodynamic
memory. Here, the slow vortex diffusion of transverse momentum in the fluid leads to a characteristic algebraic decay of
the form $\simeq B t^{-3/2}$, $B>0$ for the velocity-autocorrelation function of the
particle\cite{Alder:PRA_1:1970,Zwanzig:PRA_2:1970}. Hence, one may ask the question if the long-time tail due to
hydrodynamics dominates the long-time tail $\simeq -A t^{-5/2}$, $A>0$ due to repeated collisions of the probe particle with the
bath particles. Taking physical values from
experiments~\cite{Franosch:Nat_478:2011,Jeney:PRL_100:2008} and comparing both tails shows that, in principle, a window of time opens where
the algebraic decay due to the collision of the probe particle with bath particles dominates before hydrodynamics
becomes relevant at larger times.

Our predictions for the time-dependent response of a driven colloid can be tested, in principle, in laboratory experiments on a colloidal suspension via particle
tracking. The general scenario persists also for soft spheres and is not restricted to dilute systems. To first order in
the density, the different diffusivities of probe and bath particles can be trivially accounted for, e.g., by a rescaling 
of time. In particular, the case of a dilute and quenched array of obstacles (Lorentz model) is also included.
Simulation of the Lorentz model, where the bath particles are pinned, were added to the comparison of the mobility
and diffusivity with theory in Figs.~\ref{fig:stationary_state_mobility} and~\ref{fig:stationary_state_diffcoeff}.

Our analysis of the driven colloid shows that the nonequilibrium stationary state is inherently a nonanalytic function
of the driving such that the transport coefficients can not be expanded in a Taylor series beyond linear response.
Although this has been derived to first order in the densities only, this behavior is anticipated to be generic and
valid for arbitrary densities. Arguably, such relations should hold
universally in a general nonlinear response framework. This view is supported by recent predictions for a
two-dimensional or three-dimensional driven lattice Lorentz gas~\cite{Leitmann:PRL_111:2013,Leitmann:PRL_118:2017,Leitmann:JPhysA:2018}, where qualitatively the
same scenario applies.

\begin{acknowledgments}
We gratefully acknowledge support by the DFG research unit FOR1394 ``Nonlinear response to probe vitrification'' and
MINECO and ERDF under project FIS2015-69022P (AMP). The computational results presented have been achieved (in part)
using the HPC infrastructure LEO of the University of Innsbruck. 
\end{acknowledgments}

\appendix
\section{Computer Simulations} \label{appendix:computer_simulations}
We have simulated the motion of the pulled probe particle in the presence of bath particles interacting via a hard-core
potential using event-driven pseudo Brownian dynamics simulation~\cite{Scala:JCP_126:2007} ignoring inertial effects or hydrodynamic interactions. The starting point for the stochastic
simulation of the suspension is the Langevin equation
\begin{align} \label{eq:langevin_computer_simulation}
 \diff\vec{r}= \sqrt{2D_a} \vec{\eta}(t)\diff t + \mu F \vec{e}_z \diff t,
\end{align}
which describes the change of position $\diff\vec{r}$ of the probe particle in terms of the Gaussian white noise process
$\vec{\eta}$ with zero mean and covariance $\langle \eta_i(t)\eta_j(t)\rangle = \delta_{ij}\delta(t-t')$. For the bath
particles, the same equation with $F=0$ holds. The Langevin equation~[Eq.~\eqref{eq:langevin_computer_simulation}] is
implemented by introducing a fixed Brownian time step $\tau_B$, such that for every step, the pseudo-velocity
\begin{align}
\vec{v} = \sqrt{\frac{2 D_a}{\tau_B}}\mathcal{N}_\eta + \mu F \vec{e}_z 
\end{align}
is assigned to the probe and the bath particles ($F = 0$). Between these Brownian interrupts the particles move with constant velocities and collide elastically~\cite{Scala:JCP_126:2007}. The normal distributed random variable $\mathcal{N}_\eta$ arises from
discretization of the white noise and has zero mean and unit variance. The Brownian time step $\tau_B$ should be much smaller than
the diffusion time $\sigma^2/D_a$ and the drift time $\sigma/\mu F$.

For the colloidal case, we have used equilibrated configurations consisting of $1000$ particles at a fixed number
density $n\sigma^3 = 0.01$ and a typical value of the Brownian time step is $\tau_B \simeq 10^{-3} (\sigma^2/D_a)
/\text{max}(1, \text{Pe})$. 
For each data set, we have simulated at least $10^7$ independent trajectories.
For the case of the Lorentz system, we freeze the dynamics of the bath particles and apply the
stochastic dynamics only to the tracer particle.


%

\end{document}